\RequirePackage{lineno}
\documentclass[aps,prd,nofootinbib,twocolumn,showpacs,superscriptaddress,
preprintnumbers,floatfix]{revtex4-1}

\usepackage{amssymb}
\usepackage{amsmath}

\usepackage{graphicx}
\usepackage{natbib}
\usepackage{url}

\usepackage{psfrag}

\usepackage{xspace}

\usepackage{ifpdf}
\ifpdf
\DeclareGraphicsExtensions{.pdf, .jpg, .tif}
\usepackage[%
  pdftitle={Multiplicities of charged pions and kaons from semi-inclusive 
deep-inelastic scattering by the proton and the deuteron},%
  pdfauthor={The HERMES Collaboration},%
  pdfsubject={HERMES multiplicities},%
  pdfstartview=FitH,%
  bookmarks=true,%
  bookmarksopen=true,%
  breaklinks=true,%
  colorlinks=true,%
  linkcolor=blue,anchorcolor=blue,%
  citecolor=blue,filecolor=blue,%
  menucolor=blue,pagecolor=blue,%
  urlcolor=blue]{hyperref}
\else
\DeclareGraphicsExtensions{.eps, .jpg}
\usepackage[%
  breaklinks=true,%
  colorlinks=true,%
  linkcolor=blue,anchorcolor=blue,%
  citecolor=blue,filecolor=blue,%
  menucolor=blue,pagecolor=blue,%
  urlcolor=blue]{hyperref}
\fi

\graphicspath{{./figures/},{./figures/mults/}}

\newcommand{\xbj}{\ensuremath{x_{\mathrm{B}}}\xspace}
\newcommand{\pt}{\ensuremath{P_{h\perp}}\xspace}

\newcommand{\piz}{\ensuremath{\pi^0}\xspace}
\newcommand{\pip}{\ensuremath{\pi^+}\xspace}
\newcommand{\pim}{\ensuremath{\pi^-}\xspace}
\newcommand{\kp}{\ensuremath{\mathrm{K}^+}\xspace}
\newcommand{\km}{\ensuremath{\mathrm{K}^-}\xspace}
\newcommand{\rhoz}{\ensuremath{\rho^0}\xspace}

\begin{document}


\title{Multiplicities of charged pions and kaons from semi-inclusive 
deep-inelastic scattering by the proton and the deuteron}


\def\groupargonne{\affiliation{Physics Division, Argonne National Laboratory, Argonne, Illinois 60439-4843, USA}}
\def\groupbari{\affiliation{Istituto Nazionale di Fisica Nucleare, Sezione di Bari, 70124 Bari, Italy}}
\def\groupbeijing{\affiliation{School of Physics, Peking University, Beijing 100871, China}}
\def\groupbilbao{\affiliation{Department of Theoretical Physics, University of the Basque Country UPV/EHU, 48080 Bilbao, Spain and IKERBASQUE, Basque Foundation for Science, 48011 Bilbao, Spain}}
\def\groupcolorado{\affiliation{Nuclear Physics Laboratory, University of Colorado, Boulder, Colorado 80309-0390, USA}}
\def\groupdesy{\affiliation{DESY, 22603 Hamburg, Germany}}
\def\groupzeuthen{\affiliation{DESY, 15738 Zeuthen, Germany}}
\def\groupdubna{\affiliation{Joint Institute for Nuclear Research, 141980 Dubna,
 Russia}}
\def\grouperlangen{\affiliation{Physikalisches Institut, Universit\"at Erlangen-N\"urnberg, 91058 Erlangen, Germany}}
\def\groupferrara{\affiliation{Istituto Nazionale di Fisica Nucleare, Sezione di
 Ferrara and Dipartimento di Fisica, Universit\`a di Ferrara, 44100 Ferrara, Italy}}
\def\groupfrascati{\affiliation{Istituto Nazionale di Fisica Nucleare, Laboratori Nazionali di Frascati, 00044 Frascati, Italy}}
\def\groupgent{\affiliation{Department of Physics and Astronomy, Ghent University, 9000 Gent, Belgium}}
\def\groupgiessen{\affiliation{II. Physikalisches Institut, Justus-Liebig-Universit\"at Gie{\ss}en, 35392 Gie{\ss}en, Germany}}
\def\groupglasgow{\affiliation{SUPA, School of Physics and Astronomy, University
 of Glasgow, Glasgow G12 8QQ, United Kingdom}}
\def\groupillinois{\affiliation{Department of Physics, University of Illinois, Urbana, Illinois 61801-3080, USA}}
\def\groupmichigan{\affiliation{Randall Laboratory of Physics, University of Michigan, Ann Arbor, Michigan 48109-1040, USA }}
\def\groupmoscow{\affiliation{Lebedev Physical Institute, 117924 Moscow, Russia}}
\def\groupnikhef{\affiliation{National Institute for Subatomic Physics (Nikhef),
 1009 DB Amsterdam, The Netherlands}}
\def\groupstpetersburg{\affiliation{B. P. Konstantinov Petersburg Nuclear Physics Institute, Gatchina, 188300 Leningrad region, Russia}}
\def\groupprotvino{\affiliation{Institute for High Energy Physics, Protvino, 142281 Moscow region, Russia}}
\def\groupregensburg{\affiliation{Institut f\"ur Theoretische Physik, Universit\"at Regensburg, 93040 Regensburg, Germany}}
\def\grouprome{\affiliation{Istituto Nazionale di Fisica Nucleare, Sezione di Roma, Gruppo Collegato Sanit\`a and Istituto Superiore di Sanit\`a, 00161 Roma, Italy}}
\def\grouptriumf{\affiliation{TRIUMF, Vancouver, British Columbia V6T 2A3, Canada}}
\def\grouptokyo{\affiliation{Department of Physics, Tokyo Institute of Technology, Tokyo 152, Japan}}
\def\groupamsterdam{\affiliation{Department of Physics and Astronomy, VU University, 1081 HV Amsterdam, The Netherlands}}
\def\groupwarsaw{\affiliation{National Centre for Nuclear Research, 00-689 Warsaw, Poland}}
\def\groupyerevan{\affiliation{Yerevan Physics Institute, 375036 Yerevan, Armenia}}
\def\groupnone{\noaffiliation}


\groupargonne
\groupbari
\groupbeijing
\groupbilbao
\groupcolorado
\groupdesy
\groupzeuthen
\groupdubna
\grouperlangen
\groupferrara
\groupfrascati
\groupgent
\groupgiessen
\groupglasgow
\groupillinois
\groupmichigan
\groupmoscow
\groupnikhef
\groupstpetersburg
\groupprotvino
\groupregensburg
\grouprome
\grouptriumf
\grouptokyo
\groupamsterdam
\groupwarsaw
\groupyerevan


\author{A.~Airapetian}  \groupgiessen \groupmichigan
\author{N.~Akopov}  \groupyerevan
\author{Z.~Akopov}  \groupdesy
\author{E.C.~Aschenauer}\thanks{Now at: Brookhaven National Laboratory, Upton, New York 11772-5000, USA} \groupzeuthen 
\author{W.~Augustyniak}  \groupwarsaw
\author{R.~Avakian}  \groupyerevan
\author{A.~Avetissian}  \groupyerevan
\author{E.~Avetisyan}  \groupdesy
\author{S.~Belostotski}  \groupstpetersburg
\author{H.P.~Blok}  \groupnikhef \groupamsterdam
% ...

\collaboration{The HERMES Collaboration}



\vspace{1cm}
\begin{abstract}
	Multiplicities in semi-inclusive deep-inelastic scattering are presented for
	each charge state of $\pi^{\pm}$ and $\mathrm{K}^{\pm}$ mesons. 
	The data were collected by the HERMES experiment at the HERA
        storage ring using 27.6 GeV 
	electron and positron beams incident on a hydrogen or deuterium gas target.
	The results are presented as a function of
        the kinematic quantities \xbj, $Q^2$, $z$, and \pt.
	They represent a unique data set for identified hadrons that will
	significantly enhance our understanding of the fragmentation of quarks into
	final-state hadrons in deep-inelastic scattering.
\end{abstract}


\maketitle 

\section{Introduction}
\label{sec:intro}

Isolated quarks have never been observed in nature. When a quark
or antiquark is ejected from a bound state of quarks and gluons by the
absorption of a high-energy photon, as it separates from the ensemble, 
a shower or "jet" of hadrons
is produced.  This process may be considered to proceed through the
generation of additional quark-antiquark pairs from the color-field,
which eventually combine with the original quark or antiquark and with each other
until a configuration of color-singlet multiquark states is
reached. Understanding this hadronization process is an essential 
element of a complete picture of the
interaction of quarks in Quantum ChromoDynamics (QCD), 
and is basic to an understanding of the dynamics 
of quark-quark, gluon-gluon, and quark-gluon interactions.
It is described by the polarization-averaged fragmentation
function (FF) $D_f^h$, which is the number density of hadron type $h$
produced by the fragmentation of a 
struck quark/antiquark of flavor $f$. 

The flavor dependence of fragmentation functions provides a powerful 
tool for probing the flavor structure of the nucleon 
in hadron-induced hard scattering 
and in semi-inclusive deep-inelastic scattering (SIDIS). 
In the framework of perturbative
QCD (pQCD) at leading twist, SIDIS is 
viewed as the hard scattering of a lepton off a quark or antiquark, 
which subsequently hadronizes into, e.g.,~a final-state pion, kaon or proton
(or their antiparticles).
According to factorization theorems \cite{Collins:1996fb,Ji:2004xq},
SIDIS can be described in
leading-twist pQCD by three components: parton distribution functions
(PDFs), hard scattering cross sections, and FFs. The 
hard scattering cross section is calculable from pQCD. The
PDFs parameterize the flavor structure of the initial hadron state.
Both the PDFs and the FFs are non-perturbative quantities, but 
in the collinear framework where they are integrated over
parton transverse momentum, they are
believed to be universal, i.e.,~not to depend on the particular type of
process from which they were determined \cite{Kniehl:2000hk,Albino:2006wz}. 
The data available \cite{Arleo:2008dn} demonstrate 
that fragmentation of a quark (antiquark) 
of a specific flavor is favored for a final-state hadron that 
contains a quark (antiquark) of that flavor as a valence quark (antiquark). 
The strong flavor correlation
is reflected in the magnitude of FFs for ``favored'' and ``unfavored''
fragmentation. This correlation is exploited in SIDIS experiments
to probe the flavor structure of the nucleon through the technique of
flavor tagging, e.g., to extract 
the flavor dependence of quark-helicity
distributions in the proton and deuteron \cite{Airapetian:2004zf}.
It can be used to extract quantities of interest 
such as the flavor asymmetry in the light quark sea from meson yields 
in SIDIS on the proton and 
neutron \cite{Airapetian:2008qf,Ackerstaff:1998sr}.

While the knowledge of PDFs is highly developed, the data available to date 
for FFs have been much more limited, particularly for unfavored 
fragmentation. Most extractions of 
FFs \cite{Kretzer:2000yf,Kniehl:2000fe,Albino:2005me,Hirai:2007cx}
rely on high precision information from 
electron-positron annihilation into charged hadrons, which is
available at high energy from experiments at SLAC and 
LEP (e.g., Refs.~\cite{Buskulic:1994ft,
Abe:1998zs,Abreu:1998vq,Abbiendi:1999ry}), 
and has the advantage of not being convoluted with PDFs. However, these data do not
distinguish between quark and antiquark contributions because they are only sensitive
to the charge sum of specific hadron species (e.g., $\pi^+ +\pi^-$).
In addition, most data are taken at the mass scale of the $Z$ boson,
at which electroweak couplings become 
approximately equal and thus only 
flavor singlet combinations of FFs can be determined. Also, because
all available data is at the same energy scale, determination of the 
evolution with respect to the four-momentum transfer $Q^2$ of FFs is difficult. 
The database has been expanded
in recent years with the inclusion of results for inclusive single-hadron
production in proton-proton collisions at RHIC. These include measurements
of the transverse-momentum spectra of neutral pions at central rapidities at
 PHENIX~\cite{Adler:2003ab}, and at forward rapidities 
with STAR~\cite{Adams:2006ab},
as well as similar measurements for identified pions and kaons
for forward rapidities at BRAHMS ~\cite{Arsene:2007ab}.  

Accurate measurement
of normalized yields of specific hadrons in the final state in SIDIS, 
i.e.,~particle multiplicities,
provides a means of extracting FFs at much lower energy scales than
those of the collider measurements. The HERMES experiment with its highly
developed particle identification and pure gas targets is ideally suited
for such measurements. The data presented here were extracted from
measurements of leptoproduction  of pseudo-scalar mesons in SIDIS 
that used the 27.6 GeV lepton beam of the
HERA storage ring at DESY, which operated with
electrons or positrons. The extraction of
multiplicities of pions and kaons separately for positive
and negative charge provides sensitivity to the individual
quark and antiquark flavors in the fragmentation process. The data 
presented here for proton and deuteron targets
are the most precise results for multiplicities 
currently available at this energy scale.
With the inclusion of the kinematic dependence
of the multiplicities on the component of
hadron momentum transverse to the momentum transfer \pt, 
these data reach beyond the standard
collinear factorization, and access the transverse-momentum dependence
of the fragmentation process.
A preliminary version of a subset of these data has already been used in a recent
FF extraction \cite{Hillenbrand:2005ke,Maiheu:2006zz,deFlorian:2007aj}.
The data extend those of an earlier HERMES publication~\cite{Airapetian:2001qk} 
that reported results for pion fragmentation on the proton.

This paper is organized in the following way:
Sect.~\ref{sec:experiment} describes the experimental arrangement, while
the analysis is detailed in Sect.~\ref{sec:analysis}.
The hadron multiplicities are presented in Sect.~\ref{sec:results},
and in Sect.~\ref{sec:comparison} they are 
compared with leading order (LO) calculations based on 
recent global fits of FFs.
The results are summarized in Sec.~\ref{sec:conclusions}.

\section{The Experiment}
\label{sec:experiment}

HERMES was a fixed target experiment which used the lepton beam 
of the HERA lepton-proton collider at DESY.
The lepton ring stored electrons or positrons at an energy of 27.6 GeV.
Typical initial beam currents were between 30-50 mA.
The HERMES gaseous target \cite{Airapetian:2004yf} was  
internal to the lepton beam line.
It consisted of a 40 cm long open-ended elliptical storage cell 
aligned coaxially to the beam.
The storage cell was made of 75 $\mu$m thick Al 
and could be operated with a variety of atomic gases.
This innovative technique allowed 
the collection of data with little to no dilution from other nuclear material.
Part of the data was collected using  a polarized target
generated by an atomic beam source \cite{Nass:2003mk}, 
which could produce a jet of
polarized atomic hydrogen or deuterium with an average 
nuclear polarization around $85\%$
and an areal density of $7.6 \times 10^{13}$ $\mathrm{nucleons/cm}^2$ (hydrogen)
or $2.1 \times 10^{14}$ $\mathrm{nucleons/cm}^2$ (deuterium).
The nuclear polarization of the atoms was reversed at 1-3 minute intervals.
Unpolarized data were obtained using an unpolarized gas feed system,
operating with areal densities of $10^{14}$ to $10^{17}$ $\mathrm{nucleons/cm}^2$.
The high density unpolarized target was used only
for end-of-fill running where the beam current was much lower than the average
for the full fill.
The resulting luminosities were of the order of $10^{31}$ to 
$10^{33}$ $\mathrm{nucleons}/(\mathrm{cm}^{2}\,\mathrm{s})$. 
For this analysis, data collected with nuclear-polarized and unpolarized hydrogen 
and deuterium were used. 

\begin{figure}[t]
	\includegraphics[width=\linewidth]{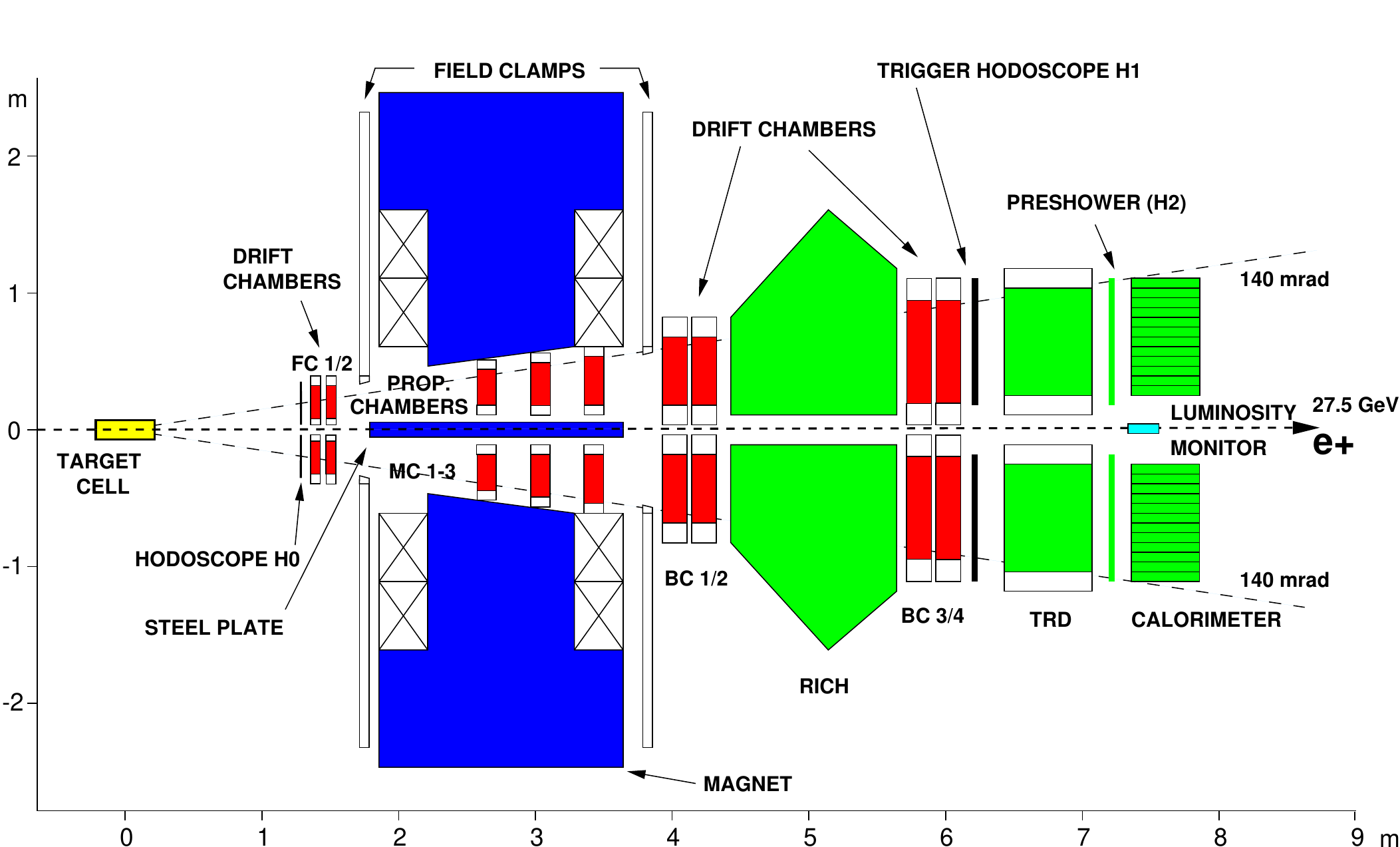}
	\caption{Schematic view of the HERMES spectrometer.}
	\label{fig:hermes}
\end{figure}

The HERMES spectrometer was a forward angle spectrometer 
with a geometrical acceptance of
$\pm 170$ mrad horizontally and $\pm | 40-140|$ mrad vertically.
The HERA beam lines passed through the spectrometer and were shielded in
the magnet area by a horizontal iron plate,
dividing the spectrometer into two symmetric halves above and below
the storage ring plane, thereby limiting the vertical acceptance
close to the beam line. The spectrometer
consisted of a front and a rear part separated by a 1.5 Tm dipole magnet.
Both parts contained tracking devices, while the back part also contained 
particle identification (PID) detectors.
A schematic side view of the spectrometer can be seen in Fig.~\ref{fig:hermes}; 
a detailed description is available in Ref.~\cite{Ackerstaff:1998av}.

Track reconstruction was performed using horizontal-drift chambers
before (FC 1/2) \cite{Brack:2001qy} and behind (BC 1/2, 3/4) \cite{Bernreuther:1998qm} the spectrometer magnet.
Combining tracks from the front and back part of the detector 
allowed the determination of the particle
momenta with an intrinsic momentum resolution $\Delta p/p$ between
0.015 and 0.025 in the accessible momentum range.

The tracks found after the spectrometer magnet 
were also used to identify hits in the PID detectors.
Lepton-hadron separation was provided by combining the information of
a Ring-Imaging \v{C}erenkov detector (RICH) \cite{Akopov:2000qi}, 
a preshower detector (H2), a
transition-radiation detector (TRD) and a lead-glass calorimeter
 \cite{Avakian:1998bz}.
More details on the use and the performance of these PID detectors
can be found in Ref.~\cite{Airapetian:2004zf}.

Charged pions, kaons, protons, and anti-protons were identified using
information from the RICH detector.
In order to achieve good separation over a momentum range of 2 to 15 GeV, the
RICH made use of two radiators.
Particles first traversed a wall of aerogel tiles with index of
refraction $n=1.03$ mounted just behind the entrance window.
The second radiator consisted of $\mathrm{C}_4\mathrm{F}_{10}$ gas
($n=1.0014$) filling the volume behind the aerogel tiles up to
a spherical-mirror array with a radius of curvature of 220 cm.
The array deflected the light produced on to a matrix of 1934 photo-multiplier
tubes per detector half.

\section{Data Analysis}
\label{sec:analysis}

The analysis was performed on data recorded in the years 2000, 2004 and 2005.
Data taken with both polarized and unpolarized targets are included, where the
regular switching of the spin
direction during polarized running results in effectively unpolarized data
when combined.
The integrated luminosities for the data sets used in this analysis are
shown in Tab.~\ref{tab:luminosities}.

The data were selected using the following criteria:
\begin{itemize}
	\item There exists a trigger composed of the signals of three hodoscopes
	(H0, H1, H2) and the electromagnetic calorimeter in 
	coincidence with the HERA lepton-bunch crossing clock. 
	\item Data quality criteria are met, including good performance of
        the particle identification and tracking detectors, and of the data 
        acquisition system.
	\item The highest momentum particle identified as a lepton
	originates from the target cell and is within the geometric acceptance 
        of the spectrometer.
        \item All other charged tracks originate from 
        the target cell and meet the geometric acceptance conditions. 
\end{itemize}
The spectrometer acceptance   
was limited by the calorimeter dimensions and
by the acceptance of the spectrometer magnet.
This translates into an acceptance in terms of the polar scattering angle 
of $40\,\mathrm{mrad} \leq \theta \leq 220\,\mathrm{mrad}$ with 
respect to the center of the target.
\begin{table}
	\begin{ruledtabular}
		\begin{tabular}{lrr}
			Year & Hydrogen              & Deuterium  \\ 
                              & (pb$^{-1}$)          & (pb$^{-1}$) \\ \hline
			2000 & 132.7 & 167.0 \\
			2004 & 31.0  & 24.3 \\
			2005 & 108.2 & 137.3
		\end{tabular}
	\end{ruledtabular}
	\caption{Integrated luminosities in $\mathrm{pb}^{-1}$ for the two target
	 gases and the three data
	taking-years included in this analysis.
	\label{tab:luminosities}}
\end{table}

\subsection{Kinematic requirements}

The kinematic constraints on events are given by the spectrometer aperture and the
requirement that the event kinematics satisfy the conditions
$Q^2>1\,\text{GeV}^2$ and $W^2>10\,\text{GeV}^2$, 
which select the deep-inelastic scattering 
(DIS) regime and suppress the region of nucleon resonances, respectively.
The relevant kinematic quantities are defined in Table~\ref{tab:kinematics}.
The limits on the fractional energy transfer to the target, $0.1 < y < 0.85$,
exclude the region where the momentum resolution starts to degrade
 \cite{Ackerstaff:1998av} (lower limit) and
the low momentum region (upper limit) where 
the trigger efficiencies have not yet reached a plateau as a function of momentum.
The upper $y$ limit also excludes the kinematic region where large radiative
corrections to inclusive cross sections are required.

\begin{table}
	\begin{ruledtabular}
		\begin{tabular}{ll}
 \parbox{0.40\columnwidth}{\raggedright $k=(E,\vec{k})$, $k' = (E', \vec{k}')$} & 
 			\parbox{0.55\columnwidth}{\raggedright  4-momenta of incident and scattered lepton $l'$}\\[4mm]
			\parbox{0.40\columnwidth}{\raggedright $P \stackrel{\mathrm{lab}}{=} (M, \vec{0})$ } & 
			\parbox{0.55\columnwidth}{\raggedright 4-momentum of the target nucleon}\\[4mm]
			\parbox{0.40\columnwidth}{\raggedright $q = k - k'$ } & 
			\parbox{0.55\columnwidth}{\raggedright 4-momentum of the virtual photon $\gamma^*$}\\[4mm]
			\parbox{0.40\columnwidth}{\raggedright $\nu = \frac{P \cdot q}{M} \stackrel{\mathrm{lab}}{=}E - E'$ } & 
			\parbox{0.55\columnwidth}{\raggedright Energy transfer to the target}\\[4mm]
			\parbox{0.45\columnwidth}{\raggedright $Q^2=-q^2\stackrel{\mathrm{lab}}{\approx}{4EE'\sin^2(\frac{\theta}{2})}$} & 
			\parbox{0.55\columnwidth}{\raggedright Negative squared four-momentum transfer}\\[4mm]
			\parbox{0.40\columnwidth}{\raggedright $W^2 = (P + q)^2$} & 
			\parbox{0.55\columnwidth}{\raggedright Squared invariant mass of the photon-nucleon system}\\[4mm]
			\parbox{0.40\columnwidth}{\raggedright $\xbj=\frac{Q^2}{2P\cdot q} \stackrel{\mathrm{lab}}{=} \frac{Q^2}{2M \cdot \nu}$ } & 
			\parbox{0.55\columnwidth}{\raggedright Bjorken scaling variable}\\[4mm]
			\parbox{0.40\columnwidth}{\raggedright $y = \frac{P \cdot q}{P \cdot k}
         \stackrel{\mathrm{lab}}{=} \frac{\nu}{E}$ } & 
			\parbox{0.55\columnwidth}{\raggedright Fractional energy of the virtual photon}\\[4mm]
			\parbox{0.40\columnwidth}{\raggedright $\phi_h$} & 
			\parbox{0.55\columnwidth}{\raggedright Azimuthal angle between the lepton scattering plane and the hadron production plane}\\[4mm]
			\parbox{0.40\columnwidth}{\raggedright $z=\frac{P \cdot P_{h}}{P \cdot q}\stackrel{\mathrm{lab}}{=} \frac{E_h}{\nu}$ } & 
			\parbox{0.55\columnwidth}{\raggedright Fractional energy of hadron $h$}\\[4mm]
			\parbox{0.40\columnwidth}{\raggedright $\pt \stackrel{\mathrm{lab}}{=} \frac{|\vec{q} \times \vec{P}_h|}{|\vec{q}|}$ } & 
			\parbox{0.55\columnwidth}{\raggedright Component of the
hadron momentum, $P_h$, transverse to $q$}\\[4mm]
		\end{tabular}
	\end{ruledtabular}
	\caption{Kinematic variables in semi-inclusive deep-inelastic scattering
	\label{tab:kinematics}}
\end{table}

Hadrons are required to have a momentum 
of $2\,\text{GeV} < P_h < 15\,\text{GeV}$,
given by the operation range of the RICH detector.
Data that are not explicitly binned in the fractional hadron energy $z$ also
are constrained to $0.2 < z < 0.8$ to exclude the region at high $z$,
which receives a sizeable contribution from exclusive processes,
and to suppress the target fragmentation region at low $z$.

\subsection{Particle identification}
\label{Sec:PID}

Leptons and hadrons are identified based on the combined responses of the 
TRD, the preshower detector, the calorimeter, and  
the RICH detector.
This response provides lepton-hadron separation with an efficiency larger 
than 98\% for leptons with contaminations $<$ 1\% and 99\% for hadrons
with a lepton contamination of $<$1\%.
The identification of charged pions and kaons using the RICH detector
is based on a direct ray tracing algorithm that
deduces the most probable particle types from 
the event-level hit pattern of
\v{C}erenkov photons on the photomultiplier matrix.
This algorithm is described in the Appendix of
~\cite{Airapetian:2012yg}.

\subsection{Multiplicities}

\begin{figure}[t]
	\includegraphics[width=\linewidth]{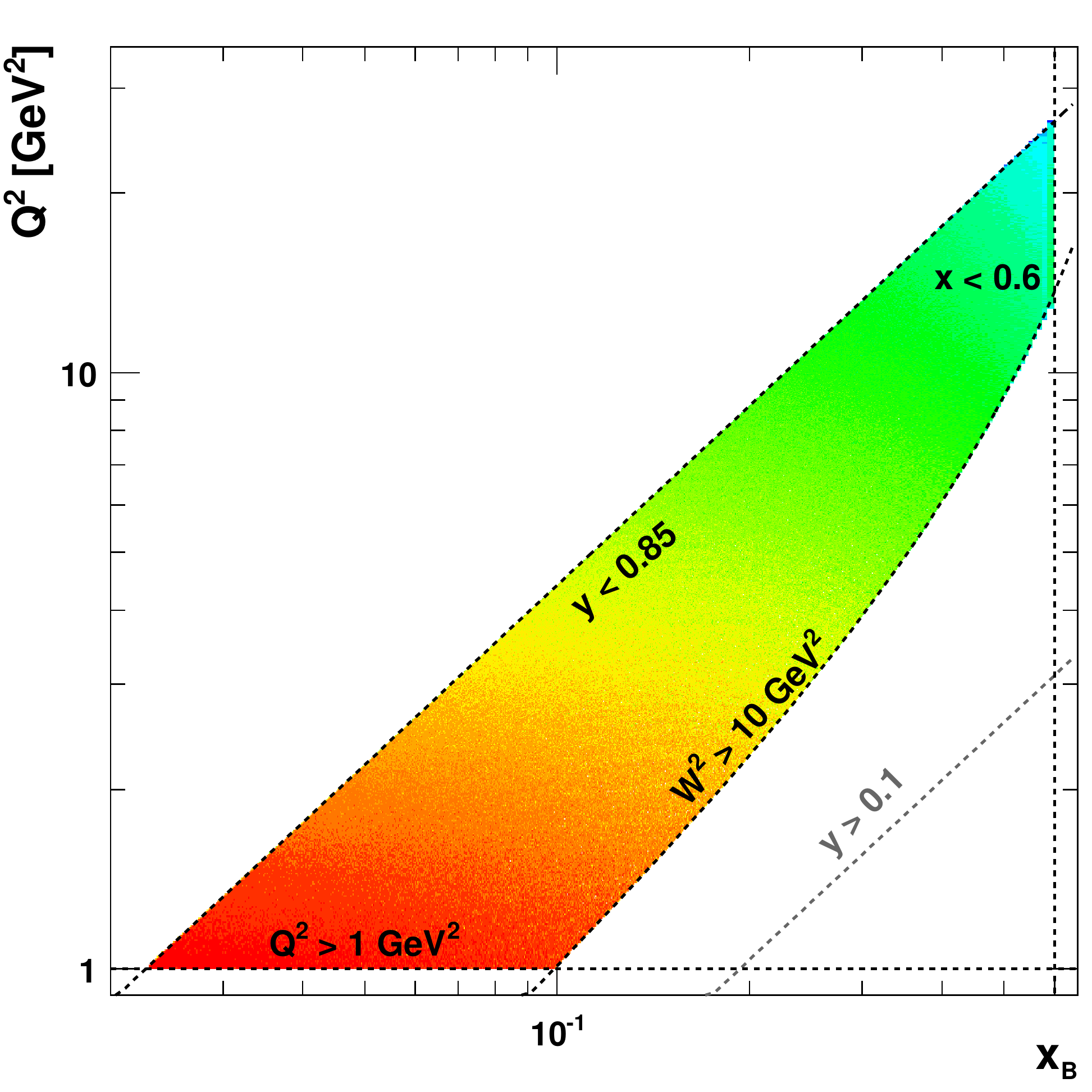}
	\caption{Born space in ($x$, $Q^{2}$) for the multiplicities extracted.}
	\label{fig:xQ2}
\end{figure}

The multiplicity ${M}_n^h$ of hadrons of the 
type $h$ produced off a target $n$ is defined 
as the respective hadron yield $N^h$ normalized
to the DIS yield.
It can be expressed in terms of the semi-inclusive cross section 
$\mathrm{d}^5\sigma^h$ and
the inclusive DIS cross section $\mathrm{d}^2\sigma_{\mathrm{DIS}}$:
\par\nobreak\noindent
\begin{eqnarray}
	& {M}_n^h (\xbj,Q^2,z,\pt)  \nonumber \\
        & = \frac{1}{\frac{\mathrm{d}^2N_{\text{DIS}}(\xbj, Q^2)}{\mathrm{d}\xbj\mathrm{d}Q^2}}
        \cdot\int_{0}^{2\pi}\frac{
	\mathrm{d}^5 N^h (\xbj, Q^2, z, \pt,\phi_h)}
	{\mathrm{d}\xbj\mathrm{d}Q^2\mathrm{d}z\,\mathrm{d}\pt
        \mathrm{d}\phi_h}\mathrm{d}\phi_h \nonumber \\
	& = 
\frac{1}{\frac{\mathrm{d}^2\sigma_{\mathrm{DIS}}(\xbj, Q^2)}{\mathrm{d}\xbj\mathrm{d}Q^2}} 
\cdot \int_{0}^{2\pi} 
\frac{\mathrm{d}^5\sigma^h(\xbj, Q^2, z, \pt, \phi_h)}{\mathrm{d}\xbj\mathrm{d}Q^2\mathrm{d}z\,\mathrm{d}\pt\,\mathrm{d}\phi_h}\mathrm{d}\phi_h.
\label{eq:mult_def}
\end{eqnarray}
The Born-level multiplicities, i.e., the multiplicities with no
limitations in geometric acceptance and corrected for radiative effects and
detector resolution, are extracted from measured
multiplicities binned in 3 dimensions: $(\xbj,z,\pt)$ when the multiplicities
as a function of $\xbj$ are desired, and $(Q^2,z,\pt)$
when they are to be given as a function of $Q^2$.
Due to the strong correlation of \xbj and $Q^2$ 
in the HERMES data (s. Fig.~\ref{fig:xQ2}) simultaneous binning in these
two variables would not have an impact on the extraction. 
A $\phi_h$ binning has been omitted because of limited statistical precision.
However, a possible impact of the $\phi_h$ dependence of the unpolarized
semi-inclusive cross section and of the acceptance 
has been accounted for in the systematic
uncertainties. The 3-dimensional
Born-level multiplicites are extracted using correction and unfolding procedures
described later in this section, which take into account charge-symmetric background, 
trigger efficiencies,  exclusive vector meson production, 
kinematic and geometric acceptance effects, and 
smearing due to radiative effects. After all corrections and unfolding
procedures are applied, the final multidimensional binned data are available, and can be
integrated over the accepted kinematic ranges (cf. Tabs.~\ref{tab:zbinning}-\ref{tab:zq2binning}) to yield the multiplicities
 in a one- or two-dimensional binning in the desired 
variables, e.g., the fractional hadron energy $z$.

\subsubsection{Charge-symmetric background}

Isolated high-energy leptons from
charge-symmetric processes like the \piz Dalitz decays and photon conversion 
into $e^+ e^-$ pairs can produce a signature indistinguishable from that of DIS events.
This background is most significant at low $Q^2$ and is much stronger
for inclusive scattering than for SIDIS.
It is taken into account by subtracting from the measured DIS or
SIDIS event rate the number of corresponding events for which the
leading lepton has a charge opposite to that of the beam particles.
The correction to the multiplicities is of the order of 1\% for DIS
and 2\% for SIDIS at low $z$.

\subsubsection{RICH unfolding}\label{Sec:RICHunf}

As described in Sec.~\ref{Sec:PID}, the hadron identification 
is based on the response of the RICH detector.
The efficiency of the detector is encoded in $3 \times 4$ matrices binned 
in momentum, charge, and event topology (number of tracks in one detector half).
They relate the vector of true hadron type $T_h$ with the 
vector of identified hadron type $I_h$:
\begin{linenomath}
\begin{equation}
	\left(
		\begin{array}{c}
			I_{\pi}        \\
			I_{\mathrm{K}} \\
			I_{\mathrm{p}} \\
			I_{\mathrm{X}}
		\end{array}
	\right) 
	= 
	\left(
		\begin{array}{ccc}
			{P}_{\pi}^{\pi} & {P}_{\mathrm{K}}^{\pi} 
& {P}_{\mathrm{p}}^{\pi} \\
			{P}_{\pi}^{\mathrm{K}} 
& {P}_{\mathrm{K}}^{\mathrm{K}} & {P}_{\mathrm{p}}^{\mathrm{K}} \\
			{P}_{\pi}^{\mathrm{P}} 
& {P}_{\mathrm{K}}^{\mathrm{p}} & {P}_{\mathrm{p}}^{\mathrm{p}} \\
			{P}_{\pi}^{\mathrm{X}} 
& {P}_{\mathrm{K}}^{\mathrm{X}} & {P}_{\mathrm{p}}^{\mathrm{X}} \\
		\end{array}
	\right)
	\cdot
		\left(		
			\begin{array}{c}
				T_{\pi} \\
				T_{\mathrm{K}} \\
				T_{\mathrm{p}}
			\end{array}
	\right).
\end{equation}
\end{linenomath}
\par\nobreak\noindent
Here, ${P}_t^i$ is the probability that a hadron of true 
type $t$ is identified as a hadron of type $i$.
The superscript $X$ refers to unidentified hadrons.
The matrices are extracted from a Monte Carlo simulation of the 
detector response that uses the PID algorithm.
Truncating the $X$ row and inverting the matrix yields a relation that gives a 
weight, the corresponding element of ${P}_{\mathrm{trunc}}^{-1}$, 
with which each identified hadron is counted as pion, kaon or proton:
\par\nobreak\noindent
\begin{equation}
	\vec{T} = {P}_{\mathrm{trunc}}^{-1} \cdot \vec{I}.
\end{equation}
The uncertainties due to RICH unfolding were estimated
to be less than 0.5\% for pions and less than 1.5\% for kaons.

\subsubsection{Trigger efficiencies}

The required trigger combines information from three hodoscopes and the
electromagnetic calorimeter.
The efficiencies of the individual detectors are extracted using special
calibration triggers, yielding an overall efficiency depending on the track
momentum and the event topology (e.g.,~events with one or two tracks)
that ranged from 95\% to 99\%.
The events are weighted with the inverted
efficiency factor.

\subsubsection{Exclusive vector-meson contribution}
\label{Sec:VMD}

Exclusive production of vector mesons ($\rhoz$, $\omega$, or $\phi$) 
can be described in the Vector Meson
Dominance (VMD) model as the fluctuation of the virtual photon into a
$q\bar{q}$-pair before its interaction with the target nucleon.
These vector mesons subsequently decay
into lighter hadrons that are then found in the final hadronic state.
The cross sections for the exclusive production show a $1/Q^6$ dependence and can be
considered as higher-twist effects.
They do not involve the fragmentation of quarks
originating from the target nucleon.
If fragmentation functions were to be extracted from multiplicities
that include such an exclusive production, they would be process dependent.
For this reason the data presented in this paper have been corrected for
hadrons stemming from these processes, but the final tabulation
includes data with and without this correction.

\begin{figure}[t]
	\centering
		\includegraphics[width=\linewidth]{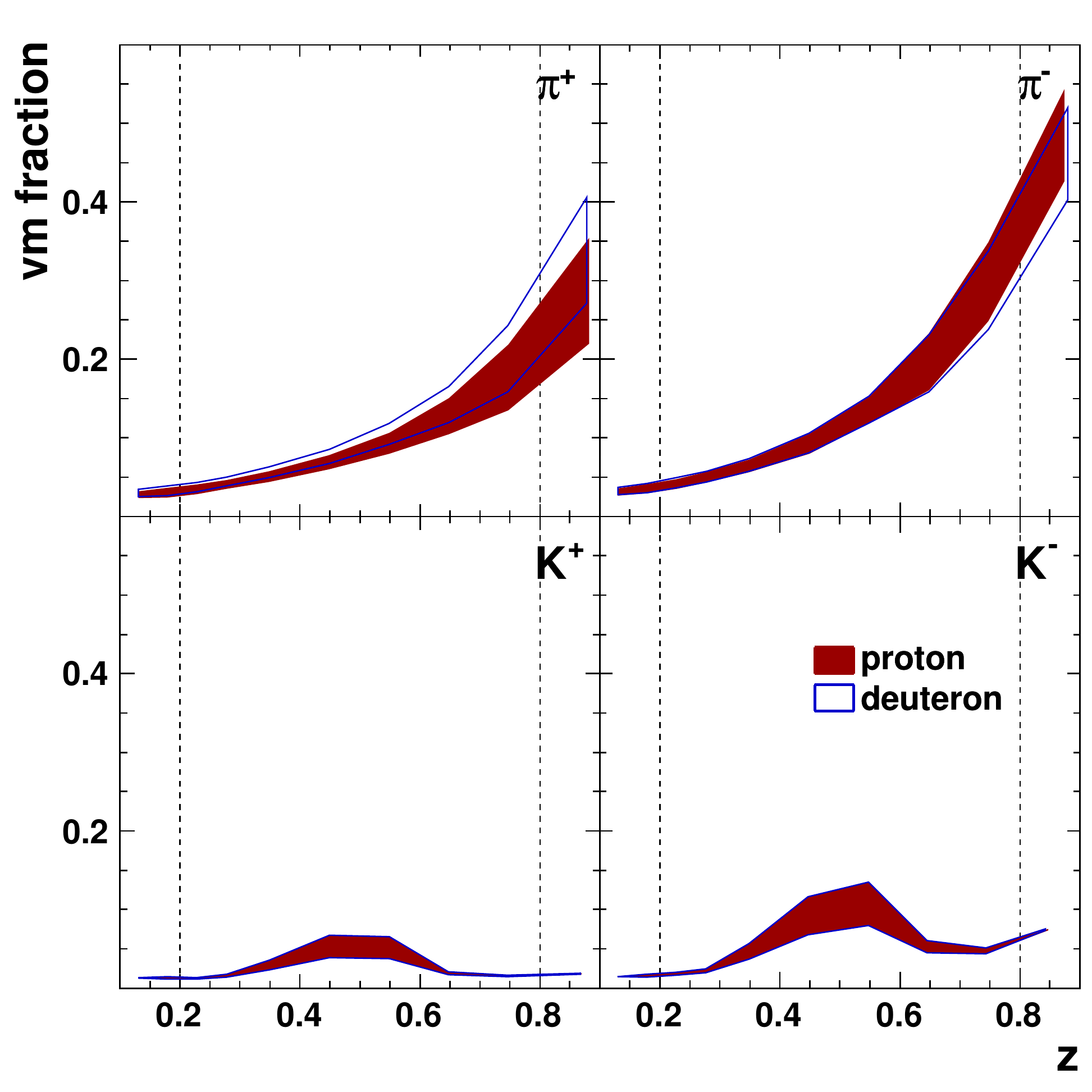}
	\caption{Fraction of mesons generated by the decay of 
        exclusive vector mesons as a function of $z$, from PYTHIA (see text).
        The widths of the bands indicate the uncertainty in the 
        corresponding fractions. The vertical dashed lines are the limits in $z$
        used in the multiplicity extractions.}
	\label{fig:vmdfrac_z}
\end{figure}

The fraction of final-state hadrons originating from exclusive
vector-meson decay was evaluated in each kinematic 
bin using the PYTHIA Monte Carlo generator.
This PYTHIA version incorporated a VMD model tuned to describe exclusive
\rhoz production at HERMES \cite{Liebing:2004us}.
Since PYTHIA can only simulate proton or neutron targets,
the values for deuterium were constructed as the combination
of the values for these nucleons.
The major contribution due to exclusive vector mesons to the final state hadron
sample arises in the form of pions originating from \rhoz decay.
Due to its anisotropic decay-angle distribution, pions from \rhoz decay
are concentrated at low and high $z$.
For the low statistics high $z$ region near $z=1$,
it is estimated that up to 50\% of the charged pions originate from \rhoz,
while at low $z$ this fraction is negligible due to the much larger yield 
from other channels. For kaons, the contribution from $\phi$ decay 
is less than 10\%. Both the SIDIS and the DIS yields
which determine the multiplicities are corrected bin by bin for exclusive vector
meson production. The correction for SIDIS is presented in its 
$z$ projection in 
Fig.~\ref{fig:vmdfrac_z}. The corresponding integrated correction for DIS is 
0.041 for the proton target and 0.044 for the deuteron target.
The bands in Fig.~\ref{fig:vmdfrac_z} for pions describe the range in 
the correction which correspond to a $1\sigma$ variation in the 
parameterization of the exclusive $\rho$ cross section. For the $\phi$
meson, the bands are half of the total variation in the correction
that results from varying the parameterization of the exclusive
$\phi$ cross section over its total range of uncertainty.

\subsubsection{Acceptance and radiative effects}

Other effects that influence the extracted multiplicities are

\begin{itemize}
	\item QED radiative effects, such as vertex corrections and the initial and final
	state radiation of real photons by the incoming
        or outgoing lepton, that alter the hard
	scattering amplitude and can mask the true kinematics at 
        the $l\gamma^*$ vertex,
	\item the limited geometric and kinematic acceptance of 
          the HERMES spectrometer,
	\item the detector resolution.
\end{itemize}

These effects were evaluated using two Monte Carlo simulations that use
the same LEPTO/JETSET \cite{Ingelman:1996mq, Sjostrand:1985ys} event generator.
The first simulation included QED radiative effects and a full detector simulation.
Radiative processes and vertex corrections were simulated 
using the RADGEN generator \cite{Akushevich:1998ft}.
The leptons and hadrons produced were tracked through a GEANT3 \cite{Brun:1978fy} 
model of the HERMES spectrometer.
Tracks were reconstructed using the same algorithm \cite{Ackerstaff:1998av} used
for real data.

As this Monte Carlo simulation contains both the generated and reconstructed
properties of any recorded track, bin to bin migration of hadron $h$
can be quantified in
a migration matrix $n^h(i,j)$, where $i$ $(=1\dots s)$ 
refers to the bin number based on reconstructed
properties and $j$ $(=0\dots s)$ to the bin number based 
on the generated (Born) properties.
The additional Born index $j=0$ refers to background events generated outside
the acceptance that migrate into it.
The probabilistic information can be summarized in a smearing matrix
\par\nobreak\noindent
\begin{equation}
	S^h(i,j) = \frac{n^h(i,j)}{n_{\mathrm{Born}}^h(j)}.
       \label{eq:matrix}
\end{equation}
The vector $n_{\mathrm{Born}}^h(j)$ is obtained from the second Monte Carlo
simulation, generated without radiative or instrumental effects.

As an example, the SIDIS yield of events without radiative or instrumental
effects, i.e., the Born yield,  ${Y}^h_{\mathrm{Born}}$ is related to the measured
SIDIS yield ${Y}^h_{\mathrm{meas}}$ via
\par\nobreak\noindent
\begin{multline}
	{Y}^h_{\mathrm{Born}}(j) = \\
	\sum_i \left[  {S'}{^h}\right]^{-1} (j,i) \left[ {Y}^h_{\mathrm{meas}}(i) 
        R - n^h(i,0)\right].
\end{multline}
Here $R=\sum_ln_{\mathrm{tracked}}^{\mathrm{DIS}}(l)/\sum_lY_{\mathrm{meas}}^{\mathrm{DIS}}(l)$
where $n^{\mathrm{DIS}}_{\mathrm{tracked}}(l)$) is the 
number of inclusive DIS events in bin $l$ in 
the MC simulation with radiative and instrumental effects and 
$Y_{\mathrm{meas}}^{\mathrm{DIS}}(i)$ is the measured number of DIS events in bin $i$.
The matrix $\left[  {S'}{^h}\right]^{-1} (j,i)$ is
the inverted SIDIS smearing matrix with $j>0$.
Because the Born multiplicity is ${Y}^h_{\mathrm{Born}}/Y_{\mathrm{Born}}^{\mathrm{DIS}}$, 
in terms of measured quantities it is given by 
\begin{multline}
	{M}^h_{\mathrm{Born}}(j) = \frac{{Y}^h_{\mathrm{Born}}(j)}
        {Y^{\mathrm{DIS}}_{\mathrm{Born}}(j)} = \\
	\frac{\sum_i \left[  {S'}{^h}\right]^{-1} (j,i) \left[ {Y}^h_{\mathrm{meas}}(i) 
        R- n^h(i,0)\right]}{\sum_k \left[  {S'}{^{\mathrm{DIS}}}\right]^{-1} (j,k) 
        \left[ {Y}^{\mathrm{DIS}}_{\mathrm{meas}}(k) R- n^{\mathrm{DIS}}(k,0)\right]}.
\end{multline}

If the kinematic binning were \emph{fully differential} in all kinematic variables,
in the limit of infinitely small bins
the smearing matrix defined by Eq.~\ref{eq:matrix} 
would be  independent of the model embodied
in the Monte Carlo event generator for $j>0$.  
If the smearing correction were applied after yields 
were integrated over some kinematic variable(s) on which both the SIDIS yields
and the spectrometer acceptance may depend, the accuracy with which the
smearing matrix corrects for acceptance effects would depend on the success
of the model in describing this yield dependence.  Hence even for those
results presented here as a function of only one kinematic variable,
this correction is applied using kinematic binning in three dimensions,
which is defined in Tabs.~\ref{tab:zbinning} to ~\ref{tab:zq2binning}.

\subsection{Systematic uncertainties}

\subsubsection{RICH unfolding}

The unfolding algorithm for correcting for the inefficiencies of the RICH detector is
based on probability matrices as described in Sec.~\ref{Sec:RICHunf}.
These probability matrices are obtained from a Monte Carlo simulation 
of the RICH detector, which produces events that are analysed with the
same particle identification algorithm (described in the Appendix of 
~\cite{Airapetian:2012yg}) as that used to analyse experimental data.
The algorithm is sensitive to the
number and distribution of background hits on the photomultiplier matrix.
Different background scenarios result in a set of slightly different probability matrices.
The variation of the final result when using the different ${P}$ 
matrices has been assigned as systematic uncertainty.
The most sensitive region is at low momentum (2-4 GeV).
Due to their lower relative flux, the uncertainty in the fraction of 
identified kaons generated by other misidentified hadrons is much 
larger than that for pions. For the multiplicities presented as a 
function of $z$, this contribution to the systematic uncertainty is 
typically less than 0.5\% for pions and 1.5\% for kaons.

\subsubsection{Azimuthal asymmetries in the unpolarized cross section}

The unpolarized cross section contains two terms generating a $\cos(\phi_h)$ and
$\cos(2\phi_h)$ modulation in the azimuthal angle $\phi_h$. 
These two terms arise, in part, from asymmetries generated by the Cahn 
\cite{Ravndal:1973kt,Kingsley:1974vf,Cahn:1978se,Cahn:1990jk} and 
the Boer-Mulders \cite{Boer:1997nt,Boer:1999si} effects.
The moments of these modulations have been 
extracted at HERMES \cite{Airapetian:2012yg}.
The effect of these modulations is not included in the Monte Carlo
simulation used for acceptance corrections. The simulation assumes an azimuthally uniform
PYTHIA6 production at the Born level. In order to study the influence of these
moments, a second extraction was carried out using a simulation in which the
events generated at the Born level are reweighted by means of a 4-dimensional
parameterization of the cosine modulations extracted from HERMES data. The difference
between the results for the extractions with the original and the reweighted
simulations was taken as the systematic uncertainty arising from the effects
of azimuthal asymmetries in the unpolarized cross section. 
For most $\xbj$ bins it is less than 2\%.

\subsubsection{Year dependence}

Both the proton and deuteron data samples are combinations of data taken
over three different years.
Comparing the results from the individual years led to differences that were
larger than expected from statistical fluctuations, which
could not be assigned to specific sources, but must be attributed to a
combination of differences in the data taking conditions
(e.g., running with polarized/unpolarized targets).
The differences are mainly concentrated in the low momentum region.
The largest deviation in the multiplicities between 
any individual data taking period and the overall result has been taken 
as a systematic uncertainty. With the exception of a few bins,
it is never larger than 2\% of the measured value.

\subsubsection{Finite bin width}

The unfolding algorithm which corrects for acceptance,
limited detector resolution and radiative effects is independent
of the Monte Carlo model used to extract the smearing
matrix if the kinematic bins are fully differential and of vanishing width.
For this case the effects of imperfections in the model
on which the SIDIS Monte Carlo is based are negligible. 
The potential impact of
any residual model dependence due to the finite bin width
is estimated by scanning over the JETSET parameters which control the Monte Carlo
event generator ~\cite{Rubin:2009zz}. This scan is 
carried out in the space in $\chi^2$ of the fit to measured
multiplicities.  
Nine parameters which constrain various features of the 
fragmentation process are varied in the scan. An eigenvector basis approach
to the Hessian method ~\cite{Pumplin:2001ct}
is used to generate nine-parameter vectors which 
are uncorrelated orthogonal combinations of the input parameters to the scan.
The intersections of these eigenvectors with the $\chi^2$ contour which lies
68\% above the best-fit minimum in the scan space, provides input
parameter sets that characterize the corresponding uncertainties in the 
multiplicities arising from those in the JETSET model parameters.  
The largest deviation of the multiplicities
from the values extracted with the standard version of the
Monte Carlo is taken as a systematic uncertainty. This
uncertainty does not exceed 3-4\%.

\section{Hadron multiplicities}
\label{sec:results}

\begin{figure}[t]
	\includegraphics[width=\linewidth]{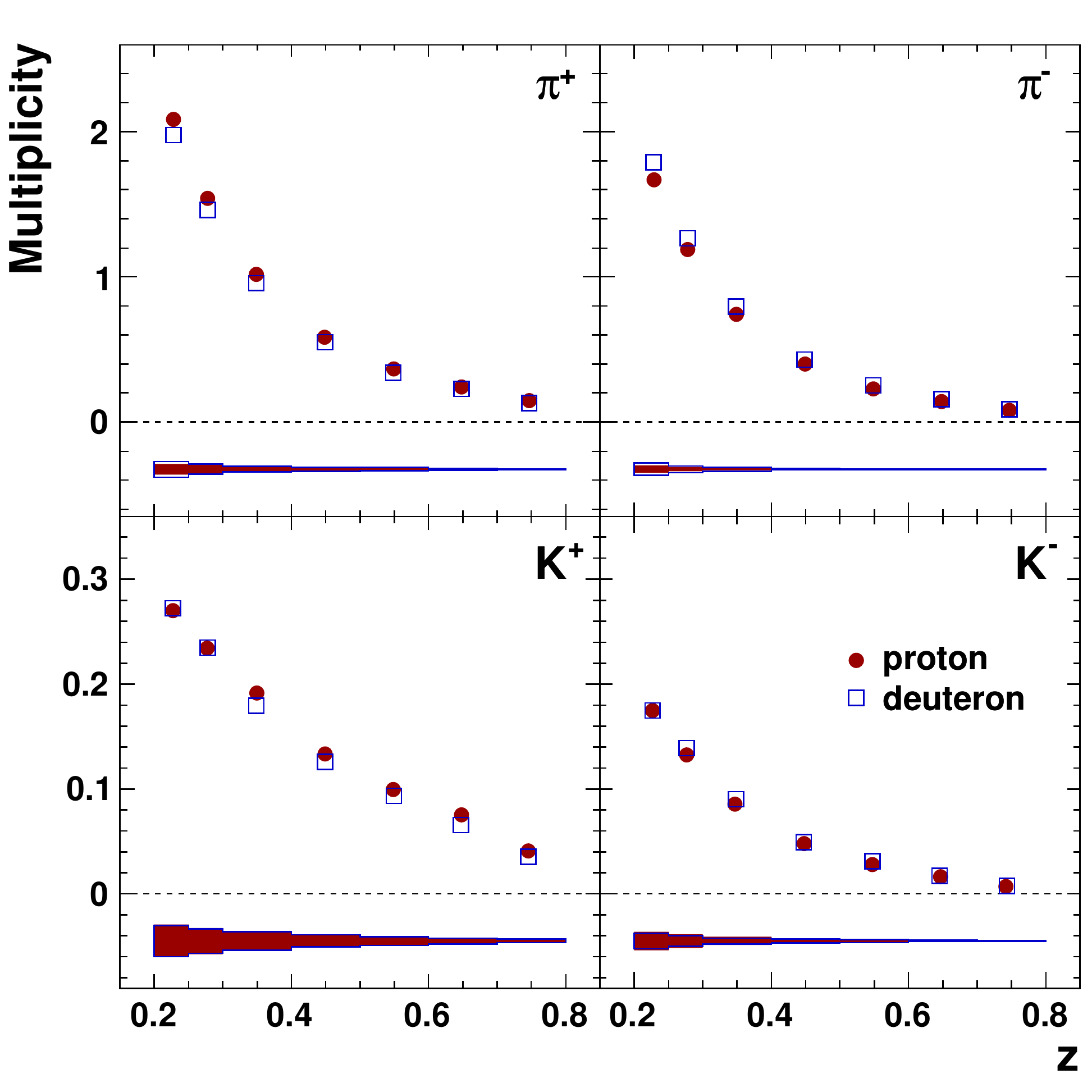}
	\caption{Multiplicities corrected for exclusive vector
        mesons as a function of $z$ from a hydrogen target (full circles) 
        and a deuterium target (empty squares). 
	Error bars for the statistical uncertainties are too 
        small to be visible.
	The systematic uncertainties are given by the error bands.
	\label{fig:mult_z}}
\end{figure}

\begin{figure*}[*t]
        \hspace{0.2cm}
        \includegraphics[width=0.41\linewidth]{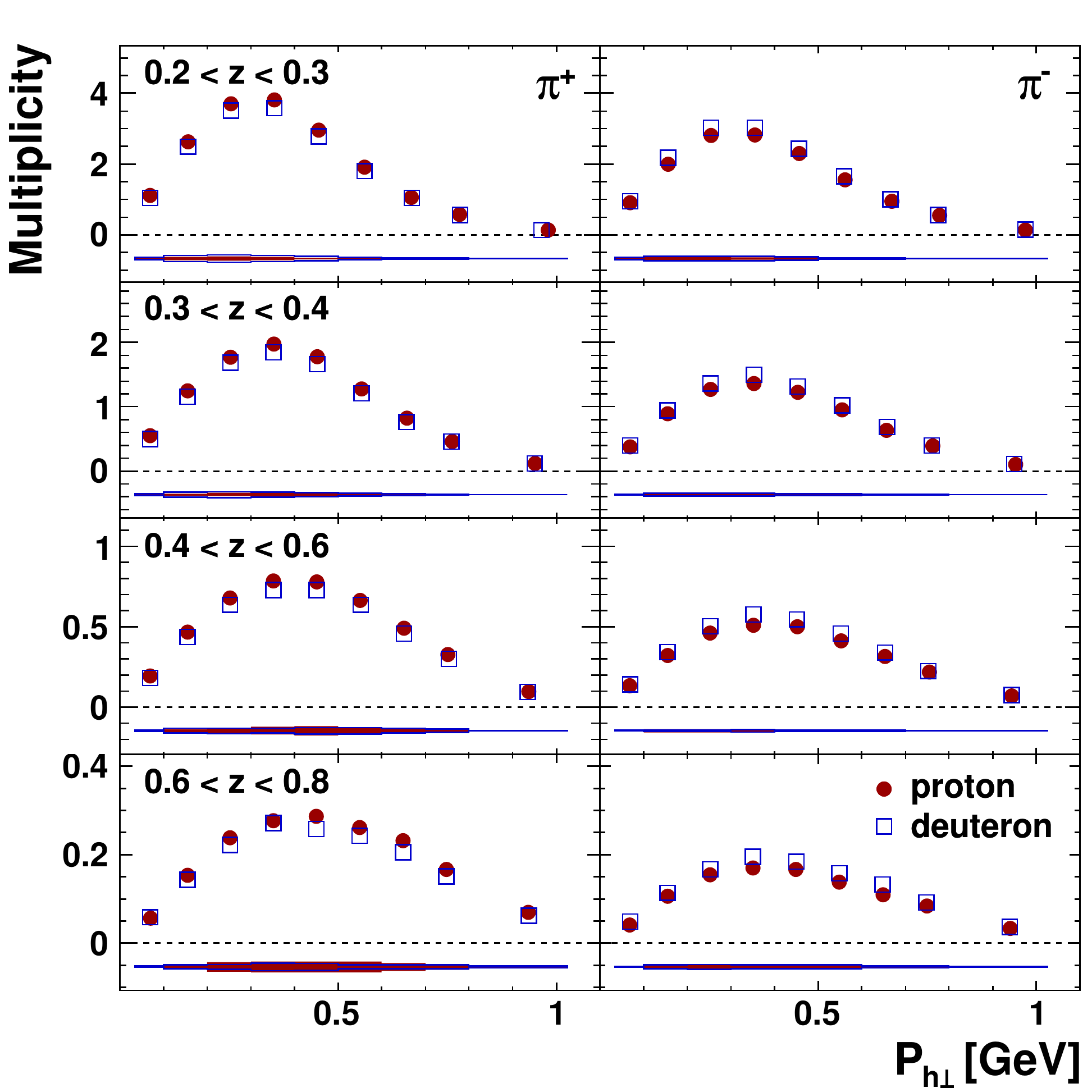}
	\includegraphics[width=0.41\linewidth]{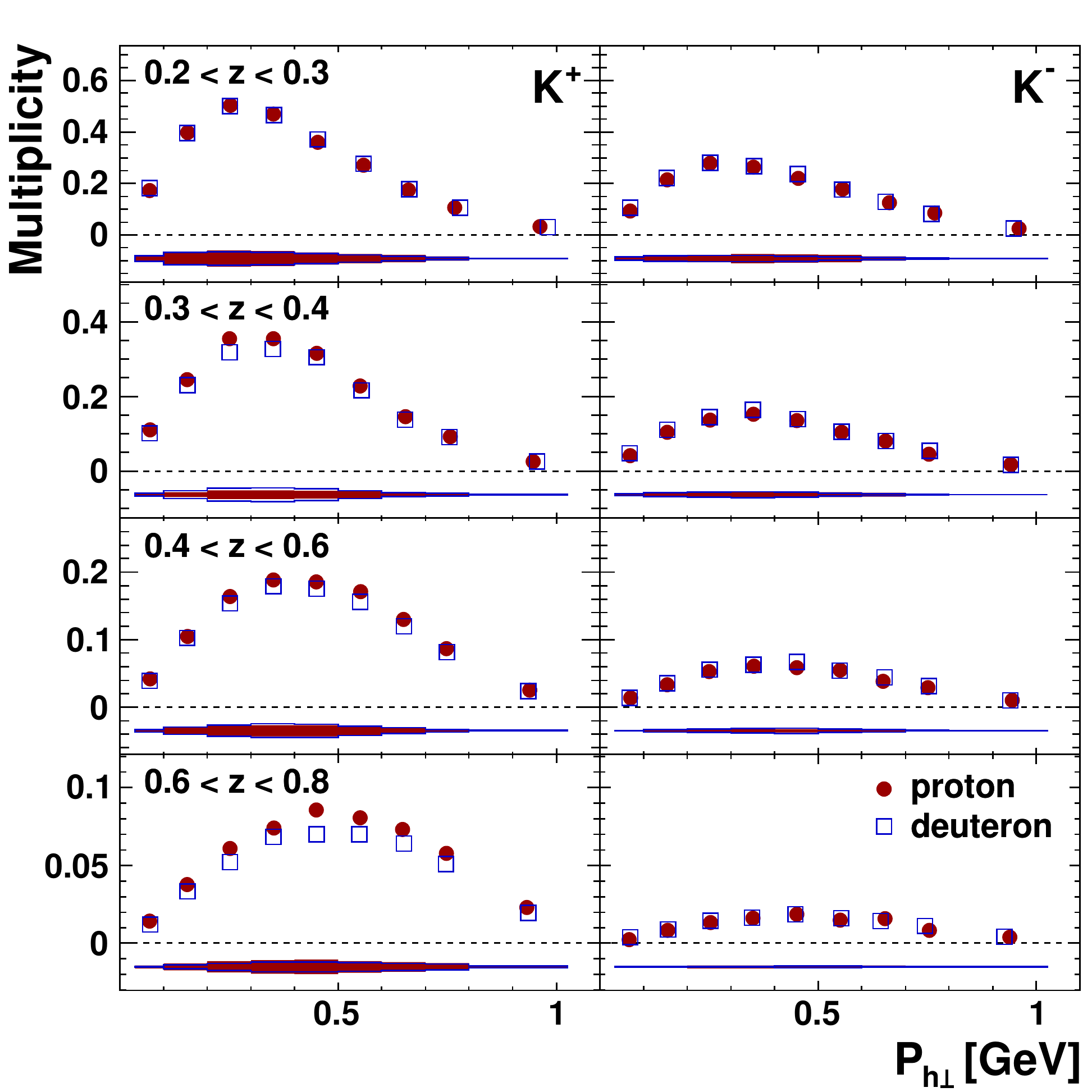} \\
        \hspace{0.2cm}
	\includegraphics[width=0.41\linewidth]{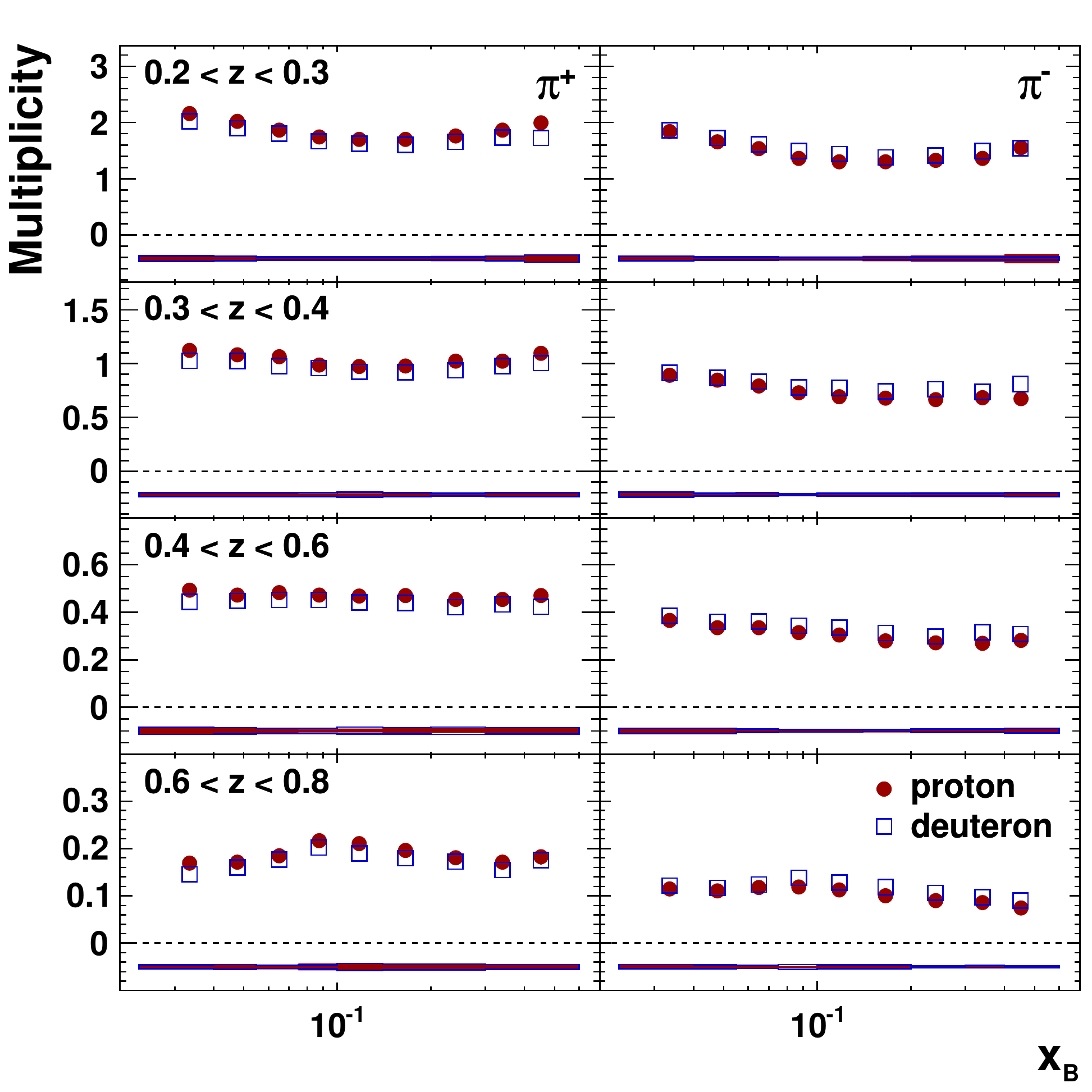}
	\includegraphics[width=0.41\linewidth]{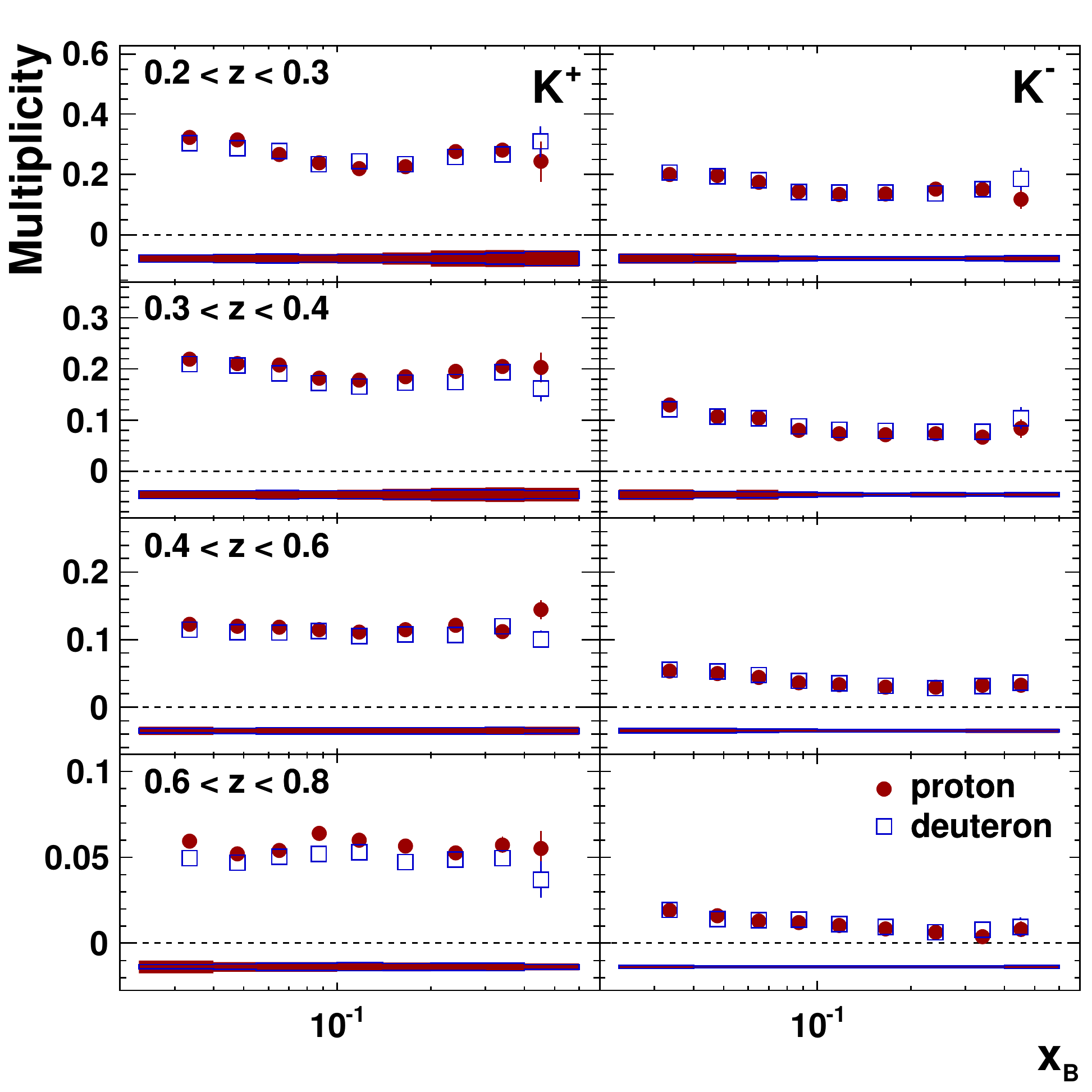} \\
        \hspace{0.2cm}
	\includegraphics[width=0.41\linewidth]{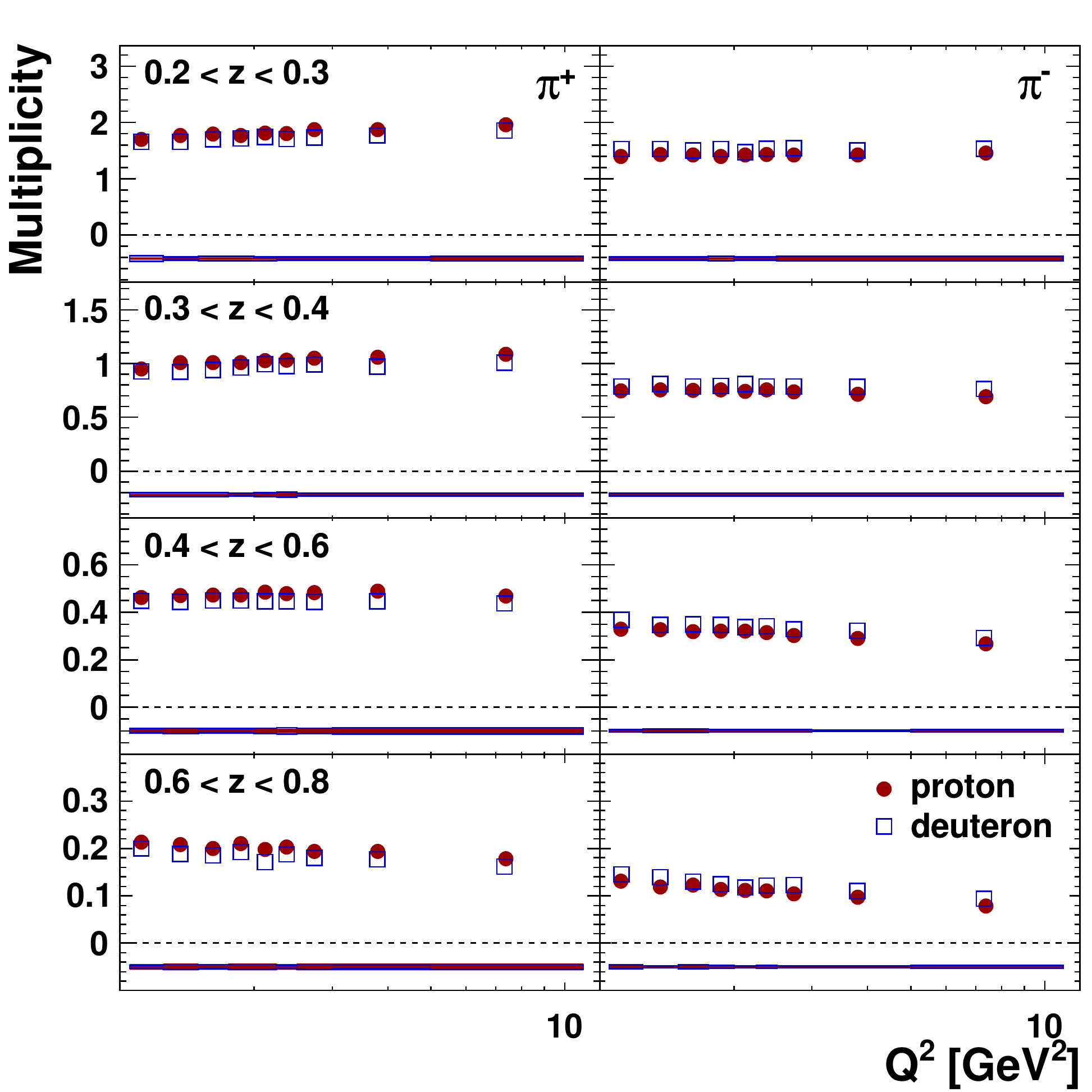}
	\includegraphics[width=0.41\linewidth]{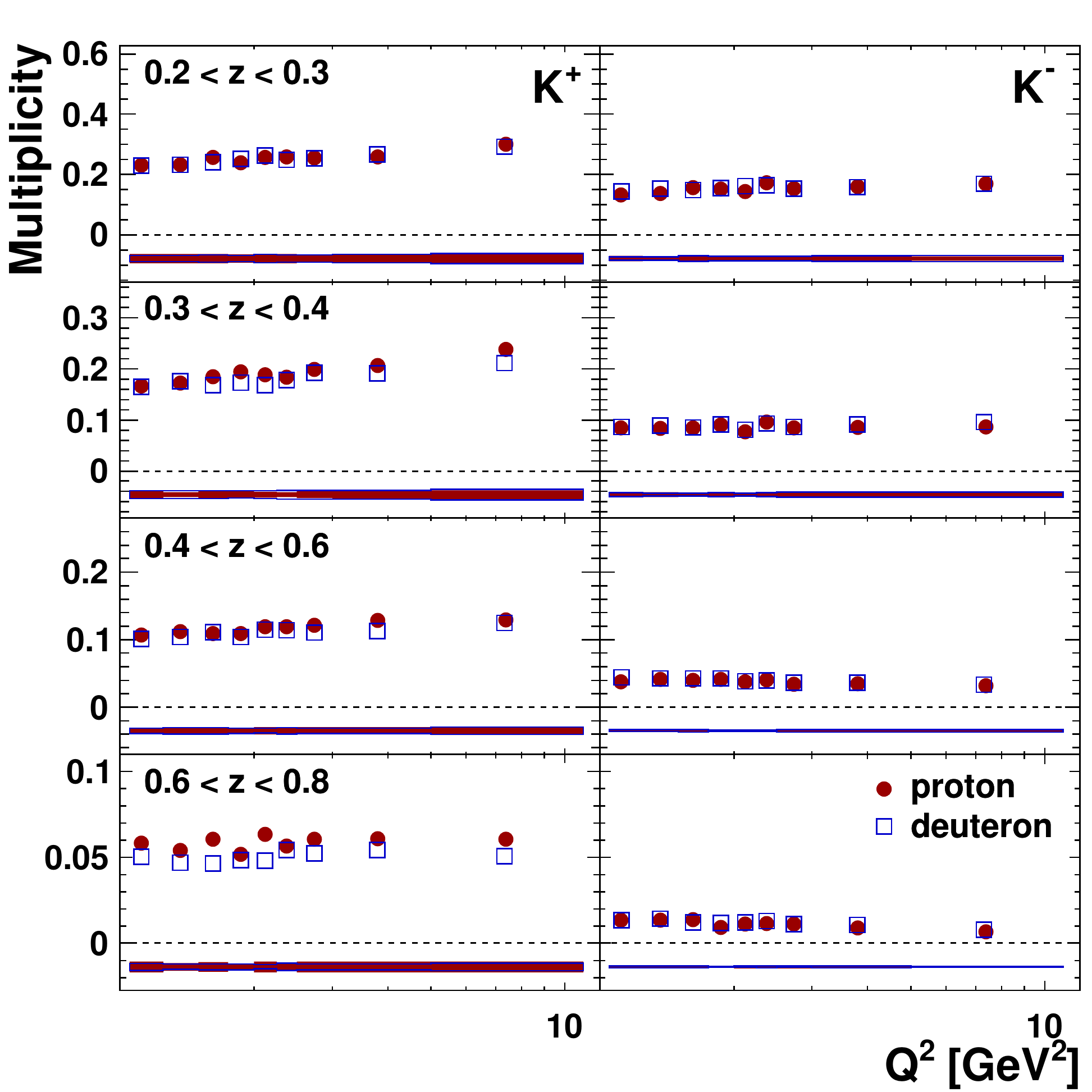}

	\caption{Multiplicities of pions (left panels) and kaons (right panels) 
        for the proton and the deuteron as a function of {\pt}, {\xbj}, 
        and $Q^2$  in four $z$ bins. Positive charge is on the left
        and negative charge is on the right of each panel. 
	Uncertainties are as in Fig.~\ref{fig:mult_z}.
	\label{fig:mult_zpt_pions}	
	}
\end{figure*}

The Born multiplicities, after correcting for exclusive-vector-meson
production, are presented in Figs.~\ref{fig:mult_z} and ~\ref{fig:mult_zpt_pions}.
In Fig.~\ref{fig:mult_z} they are shown
for both proton and deuteron targets  as a function of the
energy fraction $z$ in four panels corresponding to the type
of final-state hadron. 
\begin{figure}
	\centering
		\includegraphics[width=\linewidth]{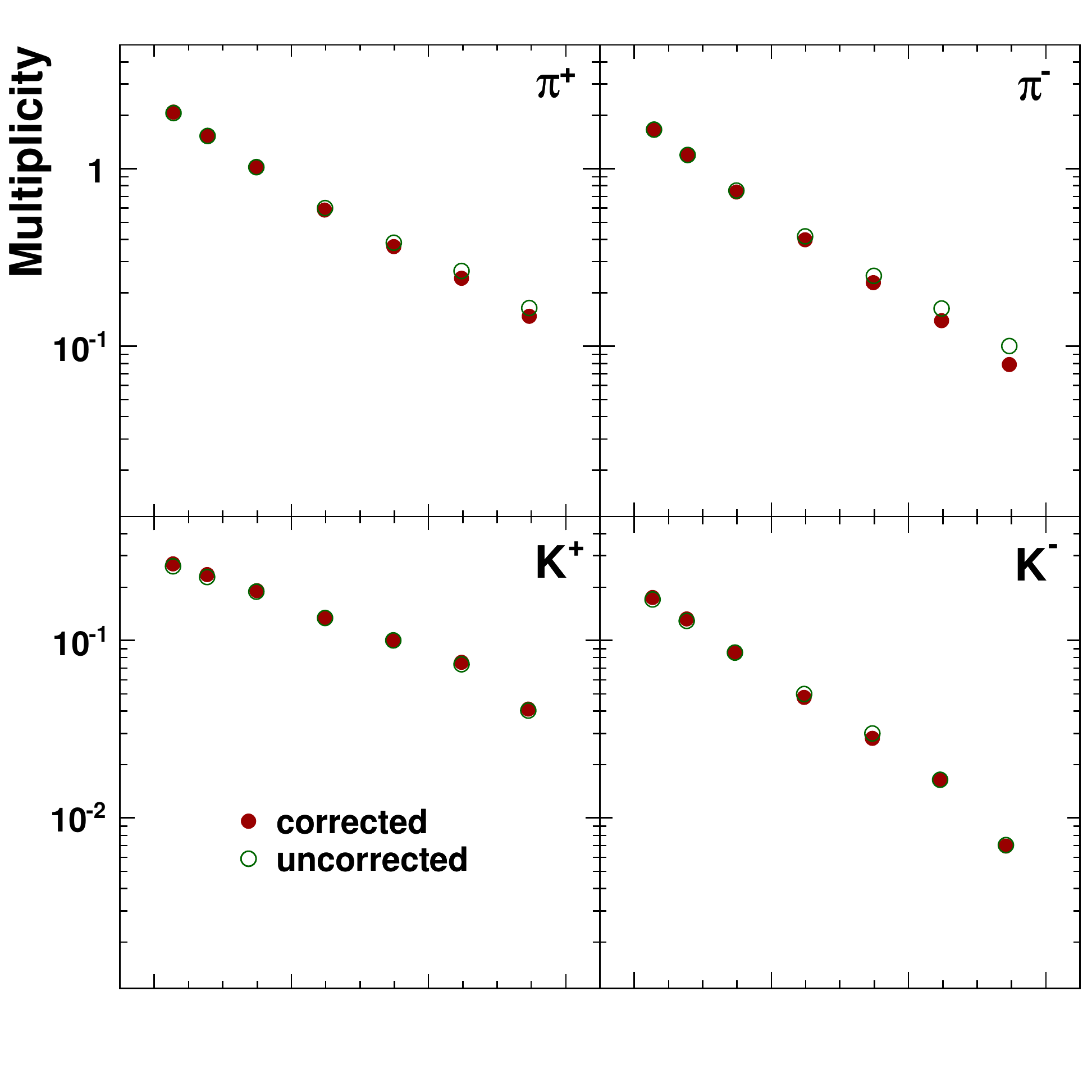}\\[-.87cm]
		\includegraphics[width=\linewidth]{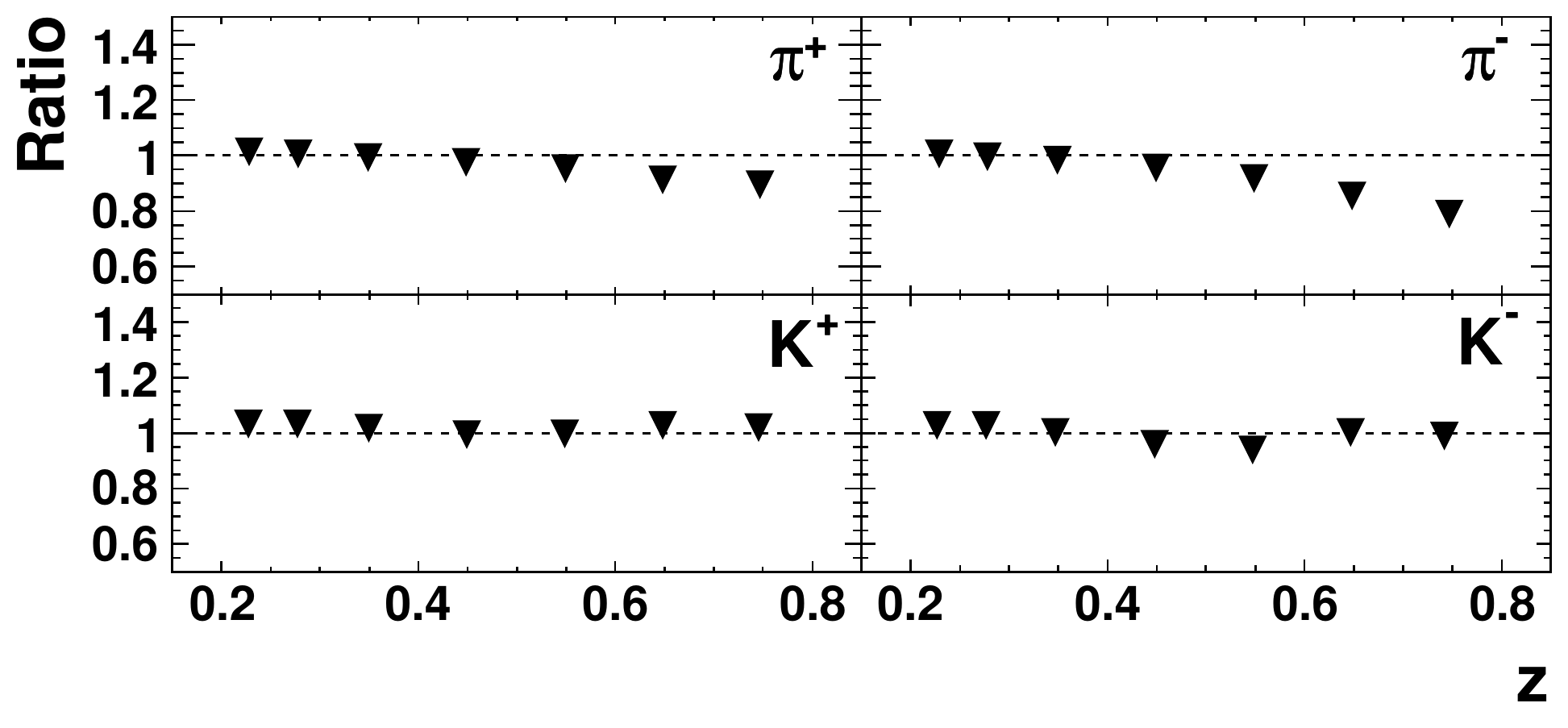}
	\caption{Comparison of measured Born multiplicities for a proton target with and without
the correction for exclusive vector mesons as a function of $z$. 
The open (closed) circles include (exclude)
exclusive vector meson production. The 
statistical error bars are too small to be visible. 
Also shown (bottom panels) are the ratios of these multiplicities. 
}
	\label{fig:mult_z_vmd}
\end{figure}
The individual panels compare data for a given hadron type 
taken with a hydrogen (full circles) or deuterium (empty squares) target.
Error bars on the points for the statistical uncertainties
are too small to be visible.
The systematic uncertainties are given by the error bands.
The $z$ bins are defined in Tab.~\ref{tab:zbinning} ($z$), which together
with Tabs.~\ref{tab:ptbinning} (\pt), \ref{tab:zxbinning} (\xbj) 
and \ref{tab:zq2binning} ($Q^2$) tabulate the binning used in the subsequent
multi-dimensional representation of the multiplicities presented 
in Fig.~\ref{fig:mult_zpt_pions}.
\begin{table}
	\begin{ruledtabular}
		\begin{tabular}{lp{7cm}}
		\xbj &
0.023 - 0.085 - 0.6 \\ \hline
    $z$ &
0.1 - 0.15 - 0.2 - 0.25 - 0.3 - 0.4 - 0.5 - 0.6 - 0.7 - 0.8 - 1.1\\ \hline
		\pt [GeV] &
0.0 - 0.1 - 0.3 - 0.45 - 0.6 - 1.2 \\
		\end{tabular}
	\end{ruledtabular}
	\caption{3D binning used for the unfolding correction of those multiplicities presented as a function of $z$ (Figs.~\ref{fig:mult_z} and \ref{fig:multdiff_z}).}
	\label{tab:zbinning}
\end{table}
To indicate the importance of the correction for exclusive vector meson decay,
the multiplicities for a proton target of pions and kaons
versus $z$ with the fraction of mesons coming from 
all processes involving exclusive vector meson decay
included (open circles) or subtracted (closed circles) 
are compared in Fig. \ref{fig:mult_z_vmd}.
Note that the highest $z$ bin in Tab.~\ref{tab:zbinning} was omitted because of the large contribution 
from exclusive vector meson production, while the two lowest $z$ bins were omitted 
because of the increased contribution due to target fragmentation. The data for these 
channels are included in the data tabulation.

The results shown in Fig. \ref{fig:mult_z} for pions (top row) from a proton 
target are in good agreement
with HERMES data published earlier~\cite{Airapetian:2001qk}. 
The proton \pip multiplicities are slightly larger than those of the deuteron, 
while for \pim the  reverse is true. 
The ratio of \pip to \pim for the proton (deuteron) 
ranges from $1.2$ ($1.1$) in the first $z$ bin 
to $2.6$ ($1.8$) in the last $z$ bin. 
These results can be attributed to the dominance of scattering off 
the $u$ quarks, reflecting the fact that the fragmentation 
process $u \rightarrow \pip$ is favored, while the 
process $u \rightarrow \pim$ is unfavored.
With rising $z$, this effect is enhanced.
Similarly, the higher $\pim$ multiplicities for the deuteron are a result
of the increased fraction of d quarks in the target and
of the stronger favored fragmentation to the $\pim$ from the neutron. 
The $\kp$ multiplicity for the proton is slightly larger
than for the deuteron, while within errors 
for $\km$ the multiplicities are equal. 
The ratio $\kp/\km$ rises from $1.5$ to $5.7$ (proton target) and from
$1.3$ to $4.6$ (deuteron target) for $0.2 < z < 0.8$ , reflecting the 
fact that \km cannot be
produced through favored fragmentation from the nucleon valence quarks.
Finally, the  $\kp/\pip$ ratio at high $z$ is about 1/3, reflecting strangeness 
suppression in fragmentation (when assuming scattering mainly from $u$ quarks).

\begin{figure}[t]
	\includegraphics[width=\linewidth]{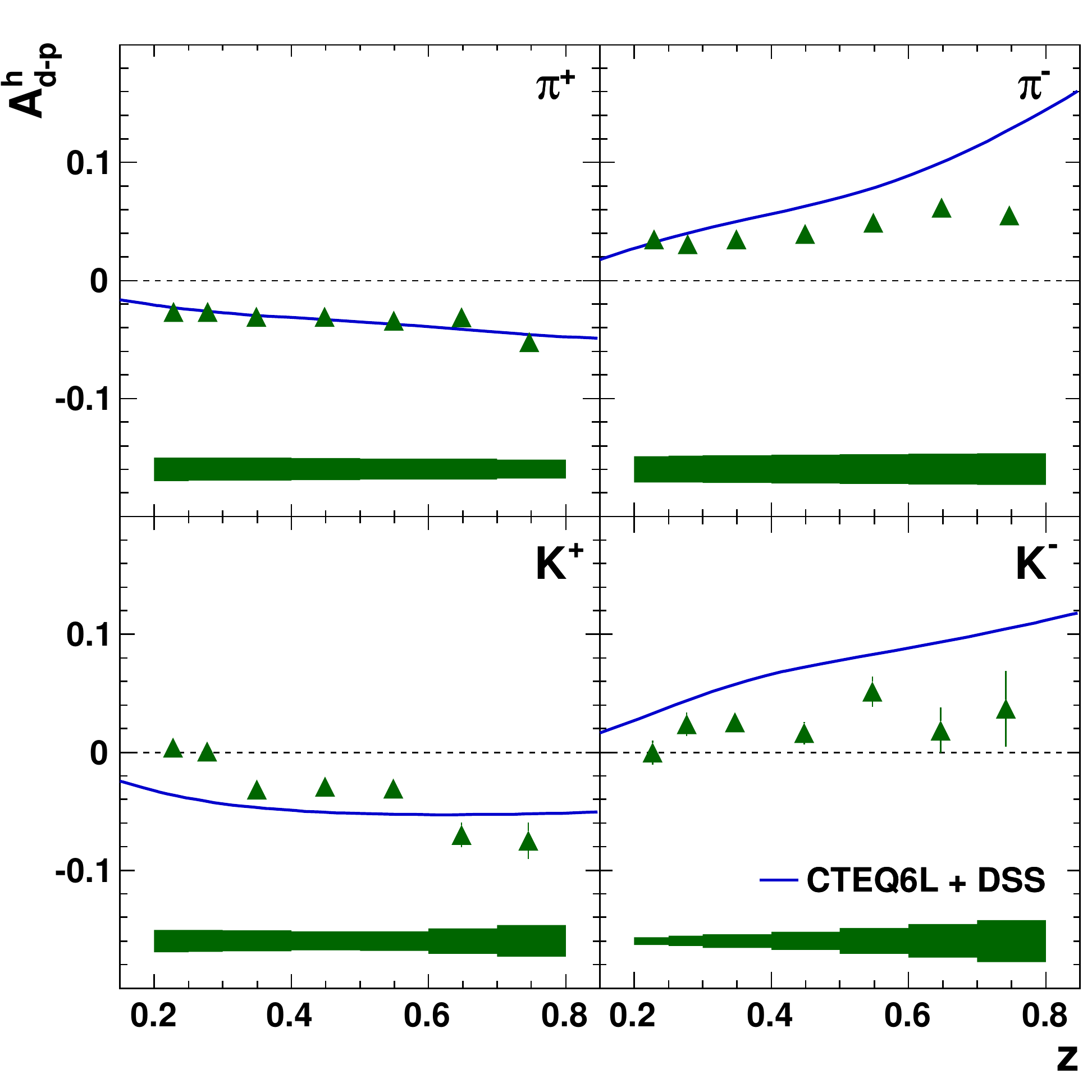}
	\caption{
	The asymmetry ${A}_{d-p}^h$ as defined in Eq.~\ref{eq:targetasy}
        as a function of $z$,
	for the multiplicities shown in Fig.~\ref{fig:mult_z}.
        The values given by an LO calculation using fragmentation functions
        from DSS~\cite{deFlorian:2007aj} and parton distributions
        from CTEQ6L~\cite{Pumplin:2002vw} are given by the solid curves.
	\label{fig:multdiff_z}}
\end{figure}

Fig. \ref{fig:multdiff_z} shows the asymmetry
\par
\begin{figure}[t]
	\includegraphics[width=\linewidth]{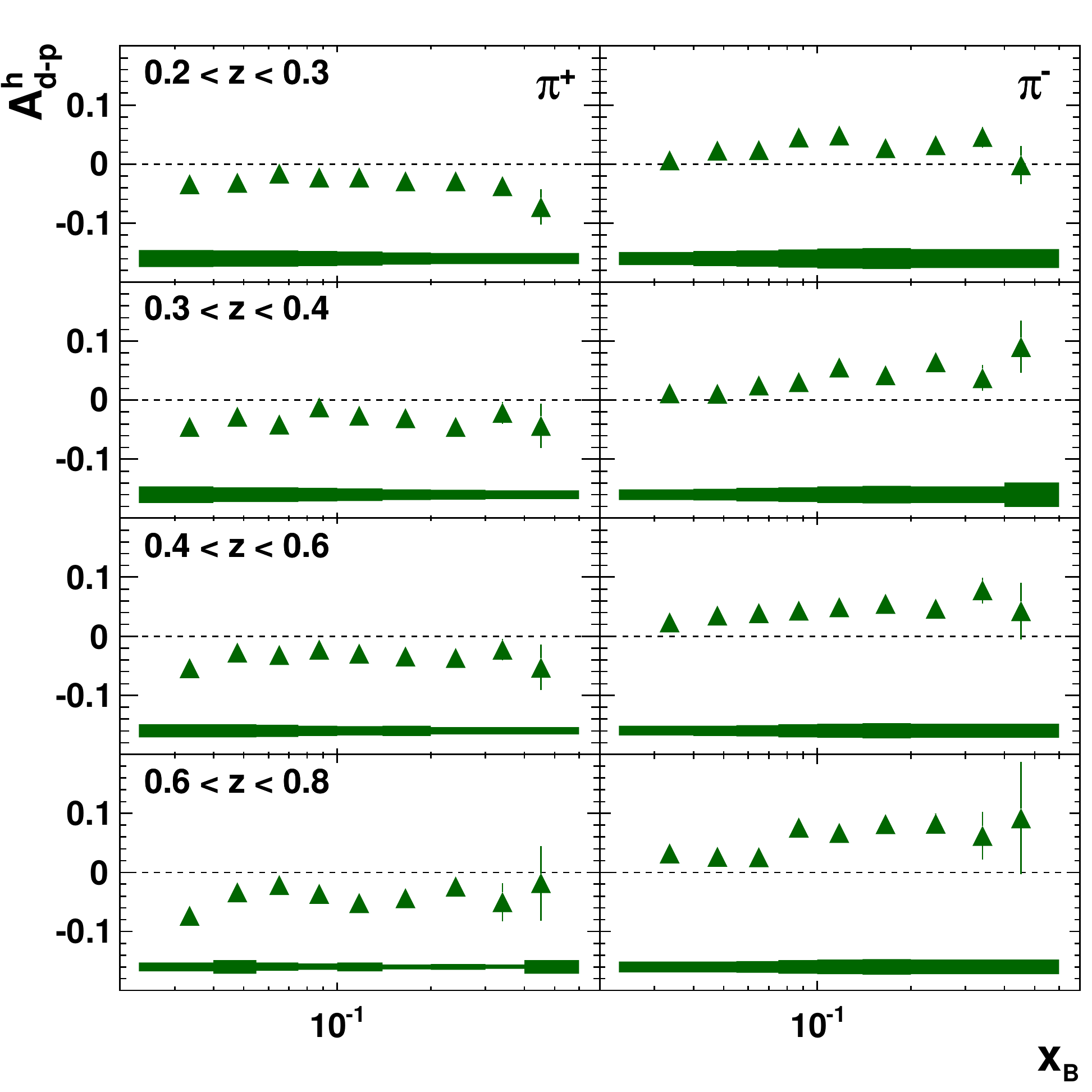}
 	\caption{
	The asymmetry ${A}_{d-p}^h$ as defined in Eq.~\ref{eq:targetasy}
        as a function of \xbj,
	for the pion multiplicities shown in Fig.~\ref{fig:mult_zpt_pions}.
	The statistical uncertainty is shown by the error bars, while the
        systematic uncertainty is given by the error bands.
	\label{fig:multdiff_zx_pions}	
	}
\end{figure}

\nobreak\noindent
\begin{equation}
	{A}^h_{{d}-{p}} = \frac{
		{M}^h_{\text{deuteron}}-{M}^h_{\text{proton}}
	}{
		{M}^h_{\text{deuteron}}+{M}^h_{\text{proton}}
	}
	\label{eq:targetasy}
\end{equation}
between the hadron production off a proton target and a deuteron target.
Because of the near equality of the proton and deuteron multiplicities
(see Fig.~\ref{fig:mult_z}), the magnitude of this asymmetry is small, but
it reflects details of the quark structure of the targets.
The negative values for \pip and the positive values for \pim reflect
the different valence quark content of the target nuclei.
The measured asymmetry in Fig. \ref{fig:multdiff_z} 
is more pronounced in the high $z$ region for kaons.
For positive kaons it is similar to that of \pip, while the corresponding
asymmetry for negative kaons is near zero, except at higher $z$, suggesting
that negative kaons are less sensitive to the valence quark content
of the target. An LO calculation (see Sect.~\ref{sec:comparison})
of $A_{d-p}^{h}$ shown in Fig.~\ref{fig:multdiff_z}
reproduces the measured values for positive charge, but strongly overpredicts
the asymmetries for negative charge. 
The same trend of negative values for \pip and the positive values for \pim
is evident in Fig.~\ref{fig:multdiff_zx_pions} where
the asymmetry $A_{d-p}^{\pi}$ is plotted as a function of \xbj
for four slices in $z$. There is no strong dependence of $A_{d-p}^{\pi}$ 
on the slice in $z$ or on \xbj. 

When the statistical precision permits, binning of extracted
multiplicities in two or more dimensions can provide useful insights into
the correlations between kinematic variables and allow for the separation
of the effects of PDFs and FFs.
In Fig.~\ref{fig:mult_zpt_pions} the multiplicities are presented
for positively and negatively charged pions and kaons
as a function of transverse hadron momentum \pt , 
\xbj , and $Q^2$ for the four slices of $z$ between 0.2
and 0.8. The features of the \pt distributions result from the combined
effects of the initial transverse motion of the struck quark in SIDIS,
of its emission of soft gluon radiation, 
and of the component of transverse momentum which is generated by 
the fragmentation process. 
The multiplicities measured at HERMES as a function of \pt 
peak at increasing \pt as $z$ is increased. 
For negative kaons the distribution versus \pt is 
wider than for positive kaons and may indicate a 
more extended fragmentation  process.
In the LUND model~\cite{Sjostrand:1993yb}, which is also
used in the HERMES Monte Carlo, fragmentation is modeled by
string fragmentation in which the color field between the quarks is the 
fragmenting entity. That field is modeled as a string with a potential energy 
that increases with increasing quark separation. When the separation
is sufficiently large the string breaks. At each string break new quark-antiquark
pairs are generated. This typically leads to more string breaks for unfavored
fragmentation, which is expected to be more relevant for negative kaons.

\begin{table}
	\begin{ruledtabular}
		\begin{tabular}{lp{7cm}}
		\xbj &
0.023 - 0.085 - 0.6 \\ \hline
    $z$ &
	   0.1 - 0.2 - 0.3 - 0.4 - 0.6 - 0.8 - 1.1 \\ \hline
		\pt [GeV] &
	 0.0 - 0.1 - 0.2 - 0.3 - 0.4 - 0.5 - 0.6 - 0.7 - 0.8 - 1.2 \\
		\end{tabular}
	\end{ruledtabular}
	\caption{3D binning used for the unfolding correction of those multiplicities 
     presented as a function of \pt and $z$ (Fig.~\ref{fig:mult_zpt_pions}).}
	\label{tab:ptbinning}
\end{table}
\begin{table}
	\begin{ruledtabular}
		\begin{tabular}{lp{7cm}}
		\xbj &
		0.023 - 0.04 - 0.055 - 0.075 - 0.1 - 
		    0.14 - 0.2 - 0.3 - 0.4 - 0.6\\ \hline
    $z$ &
	   0.1 - 0.2 - 0.3 - 0.4 - 0.6 - 0.8 - 1.1 \\ \hline
		\pt [GeV] &
		0.0 - 0.3 - 0.5 - 0.7 - 1.2\\
		\end{tabular}
	\end{ruledtabular}
	\caption{3D binning used for the unfolding correction of those multiplicities presented as a function of \xbj and $z$ (Figs.~\ref{fig:multdiff_zx_pions}
and ~\ref{fig:mult_zpt_pions}).}
	\label{tab:zxbinning}
\end{table}
\begin{table}
	\begin{ruledtabular}
		\begin{tabular}{lp{7cm}}
		$Q^2$ [GeV$^2$] &
		1. - 1.25 - 1.5 - 1.75 - 2.0 - 2.25 - 2.5 - 3.0 - 5.0 - 15.0 \\ \hline
    $z$ &
	   0.1 - 0.2 - 0.3 - 0.4 - 0.6 - 0.8 - 1.1 \\ \hline
		\pt [GeV] &
		0.0 - 0.3 - 0.5 - 0.7 - 1.2\\ 
		\end{tabular}
	\end{ruledtabular}
	\caption{3D binning used for the unfolding correction of those multiplicities presented as a function of $Q^2$ and $z$ (Fig.~\ref{fig:mult_zpt_pions}).}
	\label{tab:zq2binning}
\end{table}

In the context of the multiplicity for \pip  
and \pim being shown as a function of \xbj,
it should be noted that at HERMES kinematics 
there is a strong correlation between \xbj and $Q^2$.
For the binning given in Tab.~\ref{tab:zxbinning}, the average value
of $Q^2$ rises from 1 (GeV)$^2$ in the lowest \xbj bin to 10 (GeV)$^2$ in the
highest bin. 
A prominent feature of the multiplicity data is their kinematic dependence
on \xbj, $z$, and $Q^2$. These data are measured at a much lower energy
scale than those of most measurements of fragmentation yields.
Nevertheless, the same qualitative trends observed at
higher scales are present in the data. A strong
dependence of the multiplicities on $z$ and a weak or vanishing variation with
\xbj and $Q^2$ confirm the expectation from standard collinear factorization
 and the universality of distribution and fragmentation
functions in the SIDIS cross section at the low scales
of HERMES, $\langle Q^2 \rangle =2.5$ GeV$^2$ and $W^2\approx10$ GeV$^2$. 

A tabulation of the data is presented~\cite{DC19database} 
in four 3-dimensional decompositions corresponding
to the binnings given in Tabs.~\ref{tab:zbinning} ($z$), 
~\ref{tab:ptbinning} (\pt), \ref{tab:zxbinning} (\xbj),
and \ref{tab:zq2binning} ($Q^2$).\footnote{The data tables can also be obtained
by email to management@hermes.desy.de.}  
Because of the unfolding procedure used to extract the Born multiplicities
from the measured values, the data in the various kinematic bins are
statistically correlated. In using the data tabulated, the accompanying
covariance matrix must be considered to insure that statistical uncertainties
are not overestimated. In addition, the results of each of the projections
of these data discussed in the paper are presented.
This complete data base has been generated including both multiplicities
which have been corrected for exclusive vector meson production, as 
presented in this paper, and uncorrected multiplicities. 

An earlier extraction ~\cite{Airapetian:2008qf} of kaon multiplicities, used in an 
evaluation of the distribution of strange quarks in the nucleon, will be
superseded by data extracted using the new multiplicities 
presented here. In particular, the 
previous extraction was accomplished using one-dimensional unfolding and 
multiplicities were presented for hadron momenta larger than 2 GeV. A 
reevaluation of the strange quark distribution using the newly obtained 
more accurate kaon data will be presented in a separate 
forthcoming paper.

\section{Comparison of multiplicities with LO calculations}
\label{sec:comparison}

To date, analyses of FFs~\cite{Kretzer:2000yf,
Kniehl:2000fe,Albino:2005me,
Hirai:2007cx,deFlorian:2007aj} have been carried out in the framework of collinear factorization.
In this approximation, the multiplicity is defined as the integration of
Eq.~\ref{eq:mult_def} over \pt. In one such model, 
the LO QCD-improved quark-parton model,
the hadron multiplicity as a function of $z$ and $Q^2$ is given by
\par\nobreak\noindent
\begin{eqnarray}
  & \frac{1}{N_{DIS}(Q^2)}\frac{\mathrm{d}N^{h}(z,Q^2)}{\mathrm{d}z}
 =\frac{{\sum}_fe_f^2{\int}_0^1
q_f(\xbj ,Q^2)\mathrm{d}\xbj D_f^{\pi}(z,Q^2)}
{{\sum}_fe_f^2{\int}_0^1q_f(\xbj ,Q^2)\mathrm{d}\xbj},
\label{eq:leading}
\end{eqnarray}
where the sum is over quarks and antiquarks of flavor $f$, and $e_f$ is the 
quark charge in units of the elementary charge. The multiplicities in 
this LO approximation are a reasonable starting point  for comparing
the HERMES results with predictions based on fragmentation
functions resulting from global QCD analyses of all relevant data.

 \begin{figure}[t]
	\includegraphics[width=\linewidth]{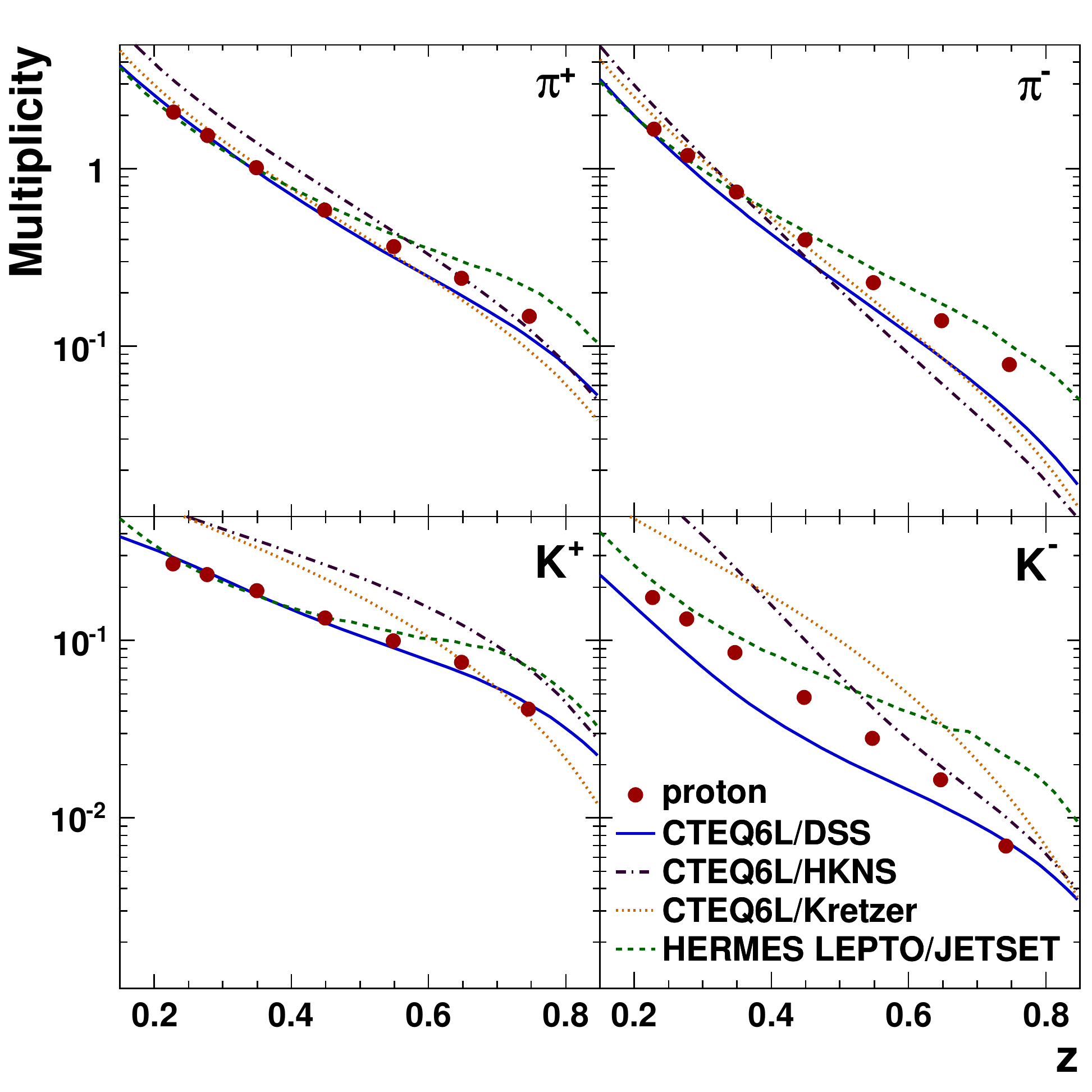}
	\caption{
	Comparison of the vector-meson-corrected 
        multiplicities measured on the proton for various hadrons 
        with LO calculations using CTEQ6L parton distributions~\cite{Pumplin:2002vw} and three
        compilations (see text) of fragmentation functions. Also shown are the
        values obtained from the HERMES Lund Monte Carlo. The statistical
        error bars on the experimental points are too small to be visible. 
  \label{fig:comparison_p}
}
\end{figure}
 \begin{figure}[t]
	\includegraphics[width=\linewidth]{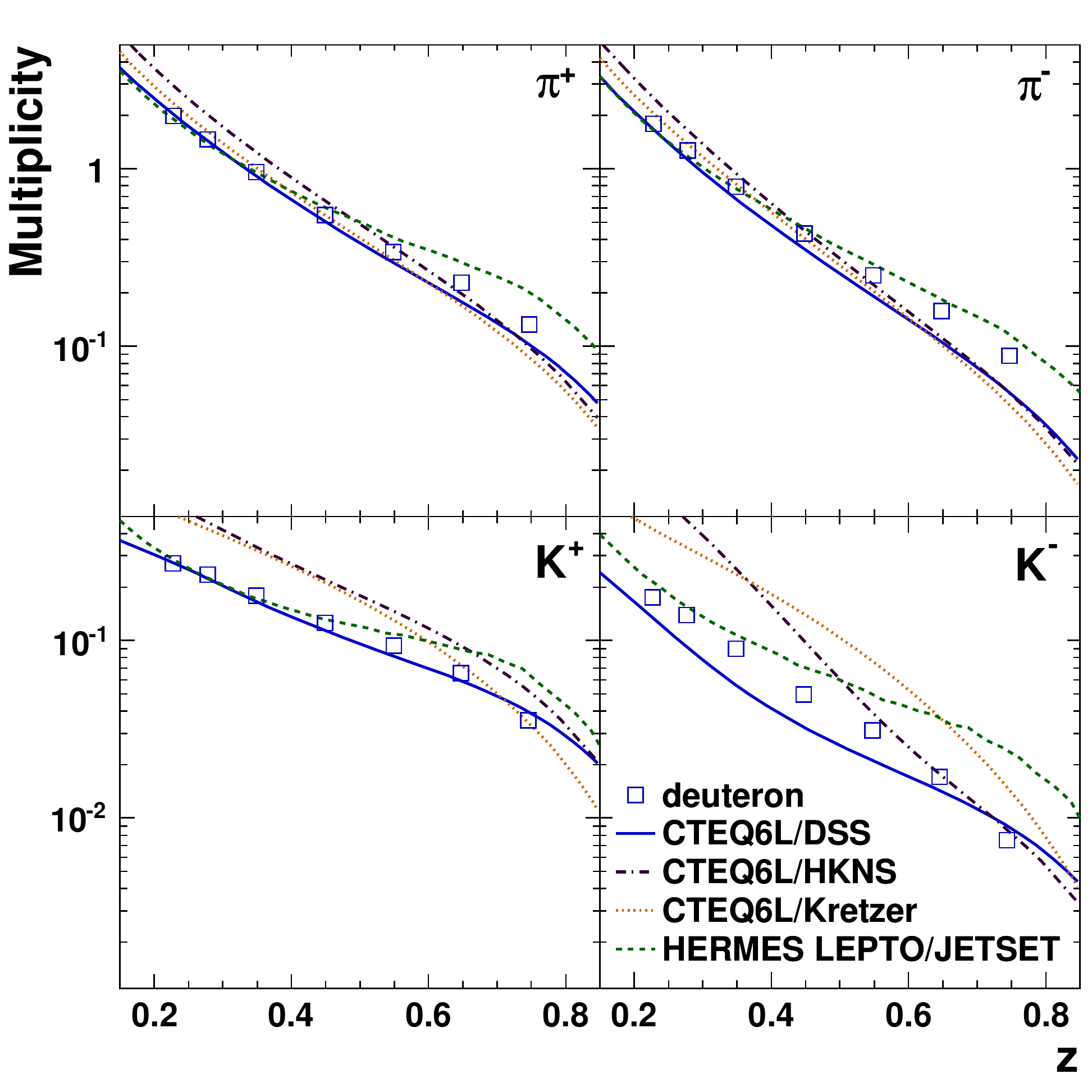}
	\caption{
	As in Fig.~\ref{fig:comparison_p} but for deuterons.
  \label{fig:comparison_d}
}
\end{figure}

A comparison of the multiplicities measured by HERMES for
SIDIS on the proton and deuteron with LO predictions is presented
in Figs.~\ref{fig:comparison_p} and \ref{fig:comparison_d}. The multiplicities are 
calculated from Eq.~\ref{eq:leading} (though integrated
only over the accepted range in \xbj of 0.023 to 0.600)
using values for the FFs taken from three widely used analyses, 
that of de Florian {\emph{et al.}} (DSS)~\cite{deFlorian:2007aj},
that of Hirai {\emph{et al.}} (HKNS)~\cite{Hirai:2007cx}, 
and that of Kretzer \cite{Kretzer:2000yf},
together with parton distributions taken from CTEQ6L~\cite{Pumplin:2002vw}. 
For positively charged pions and kaons, the results for a proton target using FFs from the 
analysis of DSS are in reasonable agreement with the HERMES results. For
negative charges, the discrepancies between data and the results
based on FFs from DSS are substantial, particularly for $K^-$ where
the curve predicted lies below the observed multiplicity 
over most of the measured range of $z$.
For $\pi^-$ the results from the DSS analysis agree with measurement at low $z$.
For both $\pi^-$ and $K^-$, fragmentation is less affected by u-quark dominance.
Uncertainties in the less abundant production by strange and anti-$u$ quarks
may have a larger impact on the predictions than for the positively charged hadrons.
Alternatively, next-to-leading-order (NLO) processes may be proportionally more
important for $\pi^-$ and particularly $K^-$, and the discrepancies observed
here may signal the importance of calculating multiplicities at NLO. 
For kaons the DSS results give a better representation
of the data than the Kretzer and HKNS curves. This is to be expected, since the DSS 
analysis included a preliminary version of the HERMES proton data in its
database. The Kretzer and HKNS results are in substantial disagreement with the 
multiplicities measured for $K^-$. The results on deuterons are in general in somewhat 
better agreement with the various predictions, in particular for pions. However,  
the discrepancy between the measured $K^{-}$ multiplicities and the various predictions is also apparent here.
In Figs.~\ref{fig:comparison_p} and \ref{fig:comparison_d}
the multiplicities obtained from the HERMES Lund Monte Carlo, in which
the fragmentation parameters have been tuned 
for HERMES kinematic conditions \cite{Hillenbrand:2005ke}, are also shown.  
Inclusion of the data reported here in future global analyses should result
in higher precision in the extraction of FFs, particularly those describing 
less abundant fragmentation processes.

\section{Summary}
\label{sec:conclusions}

HERMES has measured the multiplicity of charge-separated pions and kaons as
a function of $z$, \pt, \xbj and $Q^2$ produced by 
SIDIS off a hydrogen and a deuterium target. 
This high statistics data set, which result from scattering 
by pure gas targets of protons
and deuterons, provides unique information on the fragmentation
of quarks into final state hadrons and will contribute valuable input
for the extraction of fragmentation functions using QCD fits.
The comparison of the results from the two targets supports the notion of
fragmentation into hadrons containing the struck quark as valence quark being favored,
and should be useful in the extraction of unfavored fragmentation functions.
The multiplicities measured as a function of \pt will provide constraints on
models of the motion of quarks in the nucleon in the transverse plane 
of momentum space, as well as on the models of the fragmentation process.

\section*{Acknowledgements}

We gratefully acknowledge the DESY management for its support and the staff
at DESY and the collaborating institutions for their significant effort.
This work was supported by 
the Ministry of Economy and the Ministry of Education and Science of Armenia;
the FWO-Flanders and IWT, Belgium;
the Natural Sciences and Engineering Research Council of Canada;
the National Natural Science Foundation of China;
the Alexander von Humboldt Stiftung,
the German Bundesministerium f\"ur Bildung und Forschung (BMBF), and
the Deutsche Forschungsgemeinschaft (DFG);
the Italian Istituto Nazionale di Fisica Nucleare (INFN);
the MEXT, JSPS, and G-COE of Japan;
the Dutch Foundation for Fundamenteel Onderzoek der Materie (FOM);
the Russian Academy of Science and the Russian Federal Agency for 
Science and Innovations;
the Basque Foundation for Science (IKERBASQUE) and the UPV/EHU under program UFI 11/55;
the U.K.~Engineering and Physical Sciences Research Council, 
the Science and Technology Facilities Council,
and the Scottish Universities Physics Alliance;
the U.S.~Department of Energy (DOE) and the National Science Foundation (NSF);
as well as the European Community Research Infrastructure Integrating Activity
under the FP7 "Study of strongly interacting matter (HadronPhysics2, Grant
Agreement No. 227431)".


\bibliography{paper}

\begin{thebibliography}{47}%
\makeatletter
\providecommand \@ifxundefined [1]{%
 \@ifx{#1\undefined}
}%
\providecommand \@ifnum [1]{%
 \ifnum #1\expandafter \@firstoftwo
 \else \expandafter \@secondoftwo
 \fi
}%
\providecommand \@ifx [1]{%
 \ifx #1\expandafter \@firstoftwo
 \else \expandafter \@secondoftwo
 \fi
}%
\providecommand \natexlab [1]{#1}%
\providecommand \enquote  [1]{``#1''}%
\providecommand \bibnamefont  [1]{#1}%
\providecommand \bibfnamefont [1]{#1}%
\providecommand \citenamefont [1]{#1}%
\providecommand \href@noop [0]{\@secondoftwo}%
\providecommand \href [0]{\begingroup \@sanitize@url \@href}%
\providecommand \@href[1]{\@@startlink{#1}\@@href}%
\providecommand \@@href[1]{\endgroup#1\@@endlink}%
\providecommand \@sanitize@url [0]{\catcode `\\12\catcode `\$12\catcode
  `\&12\catcode `\#12\catcode `\^12\catcode `\_12\catcode `\%12\relax}%
\providecommand \@@startlink[1]{}%
\providecommand \@@endlink[0]{}%
\providecommand \url  [0]{\begingroup\@sanitize@url \@url }%
\providecommand \@url [1]{\endgroup\@href {#1}{\urlprefix }}%
\providecommand \urlprefix  [0]{URL }%
\providecommand \Eprint [0]{\href }%
\providecommand \doibase [0]{http://dx.doi.org/}%
\providecommand \selectlanguage [0]{\@gobble}%
\providecommand \bibinfo  [0]{\@secondoftwo}%
\providecommand \bibfield  [0]{\@secondoftwo}%
\providecommand \translation [1]{[#1]}%
\providecommand \BibitemOpen [0]{}%
\providecommand \bibitemStop [0]{}%
\providecommand \bibitemNoStop [0]{.\EOS\space}%
\providecommand \EOS [0]{\spacefactor3000\relax}%
\providecommand \BibitemShut  [1]{\csname bibitem#1\endcsname}%
\let\auto@bib@innerbib\@empty
\bibitem [{\citenamefont {Collins}\ \emph {et~al.}(1997)\citenamefont
  {Collins}, \citenamefont {Frankfurt},\ and\ \citenamefont
  {Strikman}}]{Collins:1996fb}%
  \BibitemOpen
  \bibfield  {author} {\bibinfo {author} {\bibfnamefont {J.~C.}\ \bibnamefont
  {Collins}}, \bibinfo {author} {\bibfnamefont {L.}~\bibnamefont {Frankfurt}},
  \ and\ \bibinfo {author} {\bibfnamefont {M.}~\bibnamefont {Strikman}},\
  }\href {\doibase 10.1103/PhysRevD.56.2982} {\bibfield  {journal} {\bibinfo
  {journal} {Phys. Rev.}\ }\textbf {\bibinfo {volume} {D56}},\ \bibinfo {pages}
  {2982} (\bibinfo {year} {1997})}\BibitemShut {NoStop}%
\bibitem [{\citenamefont {Ji}\ \emph {et~al.}(2004)\citenamefont {Ji},
  \citenamefont {Ma},\ and\ \citenamefont {Yuan}}]{Ji:2004xq}%
  \BibitemOpen
  \bibfield  {author} {\bibinfo {author} {\bibfnamefont {X.}~\bibnamefont
  {Ji}}, \bibinfo {author} {\bibfnamefont {J.-P.}\ \bibnamefont {Ma}}, \ and\
  \bibinfo {author} {\bibfnamefont {F.}~\bibnamefont {Yuan}},\ }\href {\doibase
  10.1016/j.physletb.2004.07.026} {\bibfield  {journal} {\bibinfo  {journal}
  {Phys. Lett.}\ }\textbf {\bibinfo {volume} {B597}},\ \bibinfo {pages} {299}
  (\bibinfo {year} {2004})}\BibitemShut {NoStop}%
\bibitem [{\citenamefont {Kniehl}\ \emph {et~al.}(2001)\citenamefont {Kniehl},
  \citenamefont {Kramer},\ and\ \citenamefont {P{\"o}tter}}]{Kniehl:2000hk}%
  \BibitemOpen
  \bibfield  {author} {\bibinfo {author} {\bibfnamefont {B.~A.}\ \bibnamefont
  {Kniehl}}, \bibinfo {author} {\bibfnamefont {G.}~\bibnamefont {Kramer}}, \
  and\ \bibinfo {author} {\bibfnamefont {B.}~\bibnamefont {P{\"o}tter}},\
  }\href
  {http://www.sciencedirect.com/science?_ob=ArticleURL&_udi=B6TVC-42H7TT5-H&_user=104183&_coverDate=03%252F12%252F2001&_rdoc=1&_fmt=&_orig=search&_sort=d&view=c&_acct=C000007298&_version=1&_urlVersion=0&_userid=104183&md5=f0b15ae8876e4542082a6fc600b812e0}
  {\bibfield  {journal} {\bibinfo  {journal} {Nucl. Phys.}\ }\textbf {\bibinfo
  {volume} {B597}},\ \bibinfo {pages} {337} (\bibinfo {year}
  {2001})}\BibitemShut {NoStop}%
\bibitem [{\citenamefont {Albino}\ \emph {et~al.}(2007)\citenamefont {Albino},
  \citenamefont {Kniehl}, \citenamefont {Kramer},\ and\ \citenamefont
  {Sandoval}}]{Albino:2006wz}%
  \BibitemOpen
  \bibfield  {author} {\bibinfo {author} {\bibfnamefont {S.}~\bibnamefont
  {Albino}}, \bibinfo {author} {\bibfnamefont {B.~A.}\ \bibnamefont {Kniehl}},
  \bibinfo {author} {\bibfnamefont {G.}~\bibnamefont {Kramer}}, \ and\ \bibinfo
  {author} {\bibfnamefont {C.}~\bibnamefont {Sandoval}},\ }\href {\doibase
  doi:10.1103/PhysRevD.75.034018} {\bibfield  {journal} {\bibinfo  {journal}
  {Phys. Rev.}\ }\textbf {\bibinfo {volume} {D75}},\ \bibinfo {pages} {034018}
  (\bibinfo {year} {2007})}\BibitemShut {NoStop}%
\bibitem [{\citenamefont {Arleo}(2009)}]{Arleo:2008dn}%
  \BibitemOpen
  \bibfield  {author} {\bibinfo {author} {\bibfnamefont {F.}~\bibnamefont
  {Arleo}},\ }\href {\doibase 10.1140/epjc/s10052-009-0871-z} {\bibfield
  {journal} {\bibinfo  {journal} {Eur. Phys. J.}\ }\textbf {\bibinfo {volume}
  {C61}},\ \bibinfo {pages} {603} (\bibinfo {year} {2009})}\BibitemShut
  {NoStop}%
\bibitem [{\citenamefont {Airapetian}\ \emph
  {et~al.}(2005{\natexlab{a}})\citenamefont {Airapetian} \emph
  {et~al.}}]{Airapetian:2004zf}%
  \BibitemOpen
  \bibfield  {author} {\bibinfo {author} {\bibfnamefont {A.}~\bibnamefont
  {Airapetian}} \emph {et~al.} (\bibinfo {collaboration} {HERMES
  Collaboration}),\ }\href {\doibase 10.1103/PhysRevD.71.012003} {\bibfield
  {journal} {\bibinfo  {journal} {Phys. Rev.}\ }\textbf {\bibinfo {volume}
  {D71}},\ \bibinfo {pages} {012003} (\bibinfo {year}
  {2005}{\natexlab{a}})}\BibitemShut {NoStop}%
\bibitem [{\citenamefont {Airapetian}\ \emph {et~al.}(2008)\citenamefont
  {Airapetian} \emph {et~al.}}]{Airapetian:2008qf}%
  \BibitemOpen
  \bibfield  {author} {\bibinfo {author} {\bibfnamefont {A.}~\bibnamefont
  {Airapetian}} \emph {et~al.} (\bibinfo {collaboration} {HERMES
  Collaboration}),\ }\href {\doibase 10.1016/j.physletb.2008.07.090} {\bibfield
   {journal} {\bibinfo  {journal} {Phys. Lett.}\ }\textbf {\bibinfo {volume}
  {B666}},\ \bibinfo {pages} {446} (\bibinfo {year} {2008})}\BibitemShut
  {NoStop}%
\bibitem [{\citenamefont {Ackerstaff}\ \emph
  {et~al.}(1998{\natexlab{a}})\citenamefont {Ackerstaff} \emph
  {et~al.}}]{Ackerstaff:1998sr}%
  \BibitemOpen
  \bibfield  {author} {\bibinfo {author} {\bibfnamefont {K.}~\bibnamefont
  {Ackerstaff}} \emph {et~al.} (\bibinfo {collaboration} {HERMES
  Collaboration}),\ }\href {\doibase 10.1103/PhysRevLett.81.5519} {\bibfield
  {journal} {\bibinfo  {journal} {Phys. Rev. Lett.}\ }\textbf {\bibinfo
  {volume} {81}},\ \bibinfo {pages} {5519} (\bibinfo {year}
  {1998}{\natexlab{a}})}\BibitemShut {NoStop}%
\bibitem [{\citenamefont {Kretzer}(2000)}]{Kretzer:2000yf}%
  \BibitemOpen
  \bibfield  {author} {\bibinfo {author} {\bibfnamefont {S.}~\bibnamefont
  {Kretzer}},\ }\href {\doibase 10.1103/PhysRevD.62.054001} {\bibfield
  {journal} {\bibinfo  {journal} {Phys. Rev.}\ }\textbf {\bibinfo {volume}
  {D62}},\ \bibinfo {pages} {054001} (\bibinfo {year} {2000})}\BibitemShut
  {NoStop}%
\bibitem [{\citenamefont {Kniehl}\ \emph {et~al.}(2000)\citenamefont {Kniehl},
  \citenamefont {Kramer},\ and\ \citenamefont {P{\"o}tter}}]{Kniehl:2000fe}%
  \BibitemOpen
  \bibfield  {author} {\bibinfo {author} {\bibfnamefont {B.~A.}\ \bibnamefont
  {Kniehl}}, \bibinfo {author} {\bibfnamefont {G.}~\bibnamefont {Kramer}}, \
  and\ \bibinfo {author} {\bibfnamefont {B.}~\bibnamefont {P{\"o}tter}},\
  }\href {\doibase 10.1016/S0550-3213(00)00303-5} {\bibfield  {journal}
  {\bibinfo  {journal} {Nucl. Phys.}\ }\textbf {\bibinfo {volume} {B582}},\
  \bibinfo {pages} {514} (\bibinfo {year} {2000})}\BibitemShut {NoStop}%
\bibitem [{\citenamefont {Albino}\ \emph {et~al.}(2005)\citenamefont {Albino},
  \citenamefont {Kniehl},\ and\ \citenamefont {Kramer}}]{Albino:2005me}%
  \BibitemOpen
  \bibfield  {author} {\bibinfo {author} {\bibfnamefont {S.}~\bibnamefont
  {Albino}}, \bibinfo {author} {\bibfnamefont {B.~A.}\ \bibnamefont {Kniehl}},
  \ and\ \bibinfo {author} {\bibfnamefont {G.}~\bibnamefont {Kramer}},\ }\href
  {\doibase 10.1016/j.nuclphysb.2005.07.010} {\bibfield  {journal} {\bibinfo
  {journal} {Nucl. Phys.}\ }\textbf {\bibinfo {volume} {B725}},\ \bibinfo
  {pages} {181} (\bibinfo {year} {2005})}\BibitemShut {NoStop}%
\bibitem [{\citenamefont {Hirai}\ \emph {et~al.}(2007)\citenamefont {Hirai},
  \citenamefont {Kumano}, \citenamefont {Nagai},\ and\ \citenamefont
  {Sudoh}}]{Hirai:2007cx}%
  \BibitemOpen
  \bibfield  {author} {\bibinfo {author} {\bibfnamefont {M.}~\bibnamefont
  {Hirai}}, \bibinfo {author} {\bibfnamefont {S.}~\bibnamefont {Kumano}},
  \bibinfo {author} {\bibfnamefont {T.-H.}\ \bibnamefont {Nagai}}, \ and\
  \bibinfo {author} {\bibfnamefont {K.}~\bibnamefont {Sudoh}},\ }\href
  {\doibase 10.1103/PhysRevD.75.094009} {\bibfield  {journal} {\bibinfo
  {journal} {Phys. Rev.}\ }\textbf {\bibinfo {volume} {D75}},\ \bibinfo {pages}
  {094009} (\bibinfo {year} {2007})}\BibitemShut {NoStop}%
\bibitem [{\citenamefont {Buskulic}\ \emph {et~al.}(1995)\citenamefont
  {Buskulic} \emph {et~al.}}]{Buskulic:1994ft}%
  \BibitemOpen
  \bibfield  {author} {\bibinfo {author} {\bibfnamefont {D.}~\bibnamefont
  {Buskulic}} \emph {et~al.},\ }\href {\doibase 10.1007/BF01556360} {\bibfield
  {journal} {\bibinfo  {journal} {Z. Phys.}\ }\textbf {\bibinfo {volume}
  {C66}},\ \bibinfo {pages} {355} (\bibinfo {year} {1995})}\BibitemShut
  {NoStop}%
\bibitem [{\citenamefont {Abe}\ \emph {et~al.}(1999)\citenamefont {Abe} \emph
  {et~al.}}]{Abe:1998zs}%
  \BibitemOpen
  \bibfield  {author} {\bibinfo {author} {\bibfnamefont {K.}~\bibnamefont
  {Abe}} \emph {et~al.},\ }\href {\doibase 10.1103/PhysRevD.59.052001}
  {\bibfield  {journal} {\bibinfo  {journal} {Phys. Rev.}\ }\textbf {\bibinfo
  {volume} {D59}},\ \bibinfo {pages} {052001} (\bibinfo {year}
  {1999})}\BibitemShut {NoStop}%
\bibitem [{\citenamefont {Abreu}\ \emph {et~al.}(1998)\citenamefont {Abreu}
  \emph {et~al.}}]{Abreu:1998vq}%
  \BibitemOpen
  \bibfield  {author} {\bibinfo {author} {\bibfnamefont {P.}~\bibnamefont
  {Abreu}} \emph {et~al.},\ }\href {\doibase 10.1007/s100529800989} {\bibfield
  {journal} {\bibinfo  {journal} {Eur. Phys. J.}\ }\textbf {\bibinfo {volume}
  {C5}},\ \bibinfo {pages} {585} (\bibinfo {year} {1998})}\BibitemShut
  {NoStop}%
\bibitem [{\citenamefont {Abbiendi}\ \emph {et~al.}(2000)\citenamefont
  {Abbiendi} \emph {et~al.}}]{Abbiendi:1999ry}%
  \BibitemOpen
  \bibfield  {author} {\bibinfo {author} {\bibfnamefont {G.}~\bibnamefont
  {Abbiendi}} \emph {et~al.} (\bibinfo {collaboration} {OPAL Collaboration}),\
  }\href {\doibase 10.1007/s100520000406} {\bibfield  {journal} {\bibinfo
  {journal} {Eur. Phys. J.}\ }\textbf {\bibinfo {volume} {C16}},\ \bibinfo
  {pages} {407} (\bibinfo {year} {2000})}\BibitemShut {NoStop}%
\bibitem [{\citenamefont {Adler}\ \emph {et~al.}(2003)\citenamefont {Adler}
  \emph {et~al.}}]{Adler:2003ab}%
  \BibitemOpen
  \bibfield  {author} {\bibinfo {author} {\bibfnamefont {S.~S.}\ \bibnamefont
  {Adler}} \emph {et~al.} (\bibinfo {collaboration} {PHENIX Collaboration}),\
  }\href {\doibase 10.1103/PhysRevLett.91.241803} {\bibfield  {journal}
  {\bibinfo  {journal} {Phys. Rev. Lett.}\ }\textbf {\bibinfo {volume} {91}},\
  \bibinfo {pages} {241803} (\bibinfo {year} {2003})}\BibitemShut {NoStop}%
\bibitem [{\citenamefont {Adams}\ \emph {et~al.}(2006)\citenamefont {Adams}
  \emph {et~al.}}]{Adams:2006ab}%
  \BibitemOpen
  \bibfield  {author} {\bibinfo {author} {\bibfnamefont {J.}~\bibnamefont
  {Adams}} \emph {et~al.} (\bibinfo {collaboration} {STAR Collaboration}),\
  }\href {\doibase 10.1103/PhysRevLett.97.152302} {\bibfield  {journal}
  {\bibinfo  {journal} {Phys. Rev. Lett.}\ }\textbf {\bibinfo {volume} {97}},\
  \bibinfo {pages} {152302} (\bibinfo {year} {2006})}\BibitemShut {NoStop}%
\bibitem [{\citenamefont {Arsene}\ \emph {et~al.}(2007)\citenamefont {Arsene}
  \emph {et~al.}}]{Arsene:2007ab}%
  \BibitemOpen
  \bibfield  {author} {\bibinfo {author} {\bibfnamefont {I.}~\bibnamefont
  {Arsene}} \emph {et~al.} (\bibinfo {collaboration} {BRAHMS Collaboration}),\
  }\href {\doibase 10.1103/PhysRevLett.98.252001} {\bibfield  {journal}
  {\bibinfo  {journal} {Phys. Rev. Lett.}\ }\textbf {\bibinfo {volume} {98}},\
  \bibinfo {pages} {252001} (\bibinfo {year} {2007})}\BibitemShut {NoStop}%
\bibitem [{\citenamefont {Hillenbrand}(2005)}]{Hillenbrand:2005ke}%
  \BibitemOpen
  \bibfield  {author} {\bibinfo {author} {\bibfnamefont {A.}~\bibnamefont
  {Hillenbrand}},\ }\href {\doibase 10.3204/DESY-THESIS-2005-035} {Ph.D.
  thesis},\ \bibinfo  {school} {Friedrich-Alexander-Universit\"at
  Erlangen-N\"urnberg} (\bibinfo {year} {2005}),\ \bibinfo {note}
  {{DESY-THESIS-2005-035}}\BibitemShut {NoStop}%
\bibitem [{\citenamefont {Maiheu}(2006)}]{Maiheu:2006zz}%
  \BibitemOpen
  \bibfield  {author} {\bibinfo {author} {\bibfnamefont {B.}~\bibnamefont
  {Maiheu}},\ }\href {http://inspirehep.net/record/740472} {Ph.D. thesis},\
  \bibinfo  {school} {Universiteit Gent} (\bibinfo {year} {2006})\BibitemShut
  {NoStop}%
\bibitem [{\citenamefont {de~Florian}\ \emph {et~al.}(2007)\citenamefont
  {de~Florian}, \citenamefont {Sassot},\ and\ \citenamefont
  {Stratmann}}]{deFlorian:2007aj}%
  \BibitemOpen
  \bibfield  {author} {\bibinfo {author} {\bibfnamefont {D.}~\bibnamefont
  {de~Florian}}, \bibinfo {author} {\bibfnamefont {R.}~\bibnamefont {Sassot}},
  \ and\ \bibinfo {author} {\bibfnamefont {M.}~\bibnamefont {Stratmann}},\
  }\href
  {http://www.google.com/search?client=safari&rls=en&q=Global+analysis+of+fragmentation+functions+for+pions+and+kaons+and+their+uncertainties&ie=UTF-8&oe=UTF-8}
  {\bibfield  {journal} {\bibinfo  {journal} {Phys. Rev.}\ }\textbf {\bibinfo
  {volume} {D75}},\ \bibinfo {pages} {114010} (\bibinfo {year}
  {2007})}\BibitemShut {NoStop}%
\bibitem [{\citenamefont {Airapetian}\ \emph {et~al.}(2001)\citenamefont
  {Airapetian} \emph {et~al.}}]{Airapetian:2001qk}%
  \BibitemOpen
  \bibfield  {author} {\bibinfo {author} {\bibfnamefont {A.}~\bibnamefont
  {Airapetian}} \emph {et~al.} (\bibinfo {collaboration} {HERMES
  Collaboration}),\ }\href {\doibase 10.1007/s100520100765} {\bibfield
  {journal} {\bibinfo  {journal} {Eur. Phys. J.}\ }\textbf {\bibinfo {volume}
  {C21}},\ \bibinfo {pages} {599} (\bibinfo {year} {2001})}\BibitemShut
  {NoStop}%
\bibitem [{\citenamefont {Airapetian}\ \emph
  {et~al.}(2005{\natexlab{b}})\citenamefont {Airapetian} \emph
  {et~al.}}]{Airapetian:2004yf}%
  \BibitemOpen
  \bibfield  {author} {\bibinfo {author} {\bibfnamefont {A.}~\bibnamefont
  {Airapetian}} \emph {et~al.} (\bibinfo {collaboration} {HERMES
  Collaboration}),\ }\href {\doibase doi:10.1016/j.nima.2004.11.020} {\bibfield
   {journal} {\bibinfo  {journal} {Nucl. Instrum. Meth.}\ }\textbf {\bibinfo
  {volume} {A540}},\ \bibinfo {pages} {68} (\bibinfo {year}
  {2005}{\natexlab{b}})}\BibitemShut {NoStop}%
\bibitem [{\citenamefont {Nass}\ \emph {et~al.}(2003)\citenamefont {Nass} \emph
  {et~al.}}]{Nass:2003mk}%
  \BibitemOpen
  \bibfield  {author} {\bibinfo {author} {\bibfnamefont {A.}~\bibnamefont
  {Nass}} \emph {et~al.},\ }\href {\doibase 10.1016/S0168-9002(03)00986-0}
  {\bibfield  {journal} {\bibinfo  {journal} {Nucl. Instrum. Meth.}\ }\textbf
  {\bibinfo {volume} {A505}},\ \bibinfo {pages} {633} (\bibinfo {year}
  {2003})}\BibitemShut {NoStop}%
\bibitem [{\citenamefont {Ackerstaff}\ \emph
  {et~al.}(1998{\natexlab{b}})\citenamefont {Ackerstaff} \emph
  {et~al.}}]{Ackerstaff:1998av}%
  \BibitemOpen
  \bibfield  {author} {\bibinfo {author} {\bibfnamefont {K.}~\bibnamefont
  {Ackerstaff}} \emph {et~al.} (\bibinfo {collaboration} {HERMES
  Collaboration}),\ }\href {\doibase 10.1016/S0168-9002(98)00769-4} {\bibfield
  {journal} {\bibinfo  {journal} {Nucl. Instrum. Meth.}\ }\textbf {\bibinfo
  {volume} {A417}},\ \bibinfo {pages} {230} (\bibinfo {year}
  {1998}{\natexlab{b}})}\BibitemShut {NoStop}%
\bibitem [{\citenamefont {Brack}\ \emph {et~al.}(2001)\citenamefont {Brack}
  \emph {et~al.}}]{Brack:2001qy}%
  \BibitemOpen
  \bibfield  {author} {\bibinfo {author} {\bibfnamefont {J.~T.}\ \bibnamefont
  {Brack}} \emph {et~al.},\ }\href {\doibase 10.1016/S0168-9002(01)00710-0}
  {\bibfield  {journal} {\bibinfo  {journal} {Nucl. Instrum. Meth.}\ }\textbf
  {\bibinfo {volume} {A469}},\ \bibinfo {pages} {47} (\bibinfo {year}
  {2001})}\BibitemShut {NoStop}%
\bibitem [{\citenamefont {Bernreuther}\ \emph {et~al.}(1998)\citenamefont
  {Bernreuther} \emph {et~al.}}]{Bernreuther:1998qm}%
  \BibitemOpen
  \bibfield  {author} {\bibinfo {author} {\bibfnamefont {S.}~\bibnamefont
  {Bernreuther}} \emph {et~al.},\ }\href {\doibase
  10.1016/S0168-9002(98)00674-3} {\bibfield  {journal} {\bibinfo  {journal}
  {Nucl. Instrum. Meth.}\ }\textbf {\bibinfo {volume} {A416}},\ \bibinfo
  {pages} {45} (\bibinfo {year} {1998})}\BibitemShut {NoStop}%
\bibitem [{\citenamefont {Akopov}\ \emph {et~al.}(2002)\citenamefont {Akopov}
  \emph {et~al.}}]{Akopov:2000qi}%
  \BibitemOpen
  \bibfield  {author} {\bibinfo {author} {\bibfnamefont {N.}~\bibnamefont
  {Akopov}} \emph {et~al.},\ }\href {\doibase 10.1016/S0168-9002(01)00932-9}
  {\bibfield  {journal} {\bibinfo  {journal} {Nucl. Instrum. Meth.}\ }\textbf
  {\bibinfo {volume} {A479}},\ \bibinfo {pages} {511} (\bibinfo {year}
  {2002})}\BibitemShut {NoStop}%
\bibitem [{\citenamefont {Avakian}\ \emph {et~al.}(1998)\citenamefont {Avakian}
  \emph {et~al.}}]{Avakian:1998bz}%
  \BibitemOpen
  \bibfield  {author} {\bibinfo {author} {\bibfnamefont {H.}~\bibnamefont
  {Avakian}} \emph {et~al.},\ }\href {\doibase 10.1016/S0168-9002(98)00540-3}
  {\bibfield  {journal} {\bibinfo  {journal} {Nucl. Instrum. Meth.}\ }\textbf
  {\bibinfo {volume} {A417}},\ \bibinfo {pages} {69} (\bibinfo {year}
  {1998})}\BibitemShut {NoStop}%
\bibitem [{\citenamefont {Airapetian}\ \emph {et~al.}(2013)\citenamefont
  {Airapetian} \emph {et~al.}}]{Airapetian:2012yg}%
  \BibitemOpen
  \bibfield  {author} {\bibinfo {author} {\bibfnamefont {A.}~\bibnamefont
  {Airapetian}} \emph {et~al.} (\bibinfo {collaboration} {HERMES
  Collaboration}),\ }\href {\doibase 10.1103/PhysRevD.87.012010} {\bibfield
  {journal} {\bibinfo  {journal} {Phys. Rev.}\ }\textbf {\bibinfo {volume}
  {D87}},\ \bibinfo {pages} {012010} (\bibinfo {year} {2013})}\BibitemShut
  {NoStop}%
\bibitem [{\citenamefont {Liebing}(2004)}]{Liebing:2004us}%
  \BibitemOpen
  \bibfield  {author} {\bibinfo {author} {\bibfnamefont {P.}~\bibnamefont
  {Liebing}},\ }\href {\doibase 10.3204/DESY-THESIS-2004-036} {Ph.D. thesis},\
  \bibinfo  {school} {Universit\"at Hamburg} (\bibinfo {year} {2004}),\
  \bibinfo {note} {{DESY-THESIS-2004-036}}\BibitemShut {NoStop}%
\bibitem [{\citenamefont {Ingelman}\ \emph {et~al.}(1997)\citenamefont
  {Ingelman}, \citenamefont {Edin},\ and\ \citenamefont
  {Rathsman}}]{Ingelman:1996mq}%
  \BibitemOpen
  \bibfield  {author} {\bibinfo {author} {\bibfnamefont {G.}~\bibnamefont
  {Ingelman}}, \bibinfo {author} {\bibfnamefont {A.}~\bibnamefont {Edin}}, \
  and\ \bibinfo {author} {\bibfnamefont {J.}~\bibnamefont {Rathsman}},\ }\href
  {\doibase 10.1016/S0010-4655(96)00157-9} {\bibfield  {journal} {\bibinfo
  {journal} {Comput. Phys. Commun.}\ }\textbf {\bibinfo {volume} {101}},\
  \bibinfo {pages} {108} (\bibinfo {year} {1997})}\BibitemShut {NoStop}%
\bibitem [{\citenamefont {Sj\"{o}strand}(1986)}]{Sjostrand:1985ys}%
  \BibitemOpen
  \bibfield  {author} {\bibinfo {author} {\bibfnamefont {T.}~\bibnamefont
  {Sj\"{o}strand}},\ }\href {\doibase 10.1016/0010-4655(86)90096-2} {\bibfield
  {journal} {\bibinfo  {journal} {Comput. Phys. Commun.}\ }\textbf {\bibinfo
  {volume} {39}},\ \bibinfo {pages} {347} (\bibinfo {year} {1986})}\BibitemShut
  {NoStop}%
\bibitem [{\citenamefont {Akushevich}\ \emph {et~al.}(1998)\citenamefont
  {Akushevich}, \citenamefont {B\"{o}ttcher},\ and\ \citenamefont
  {Ryckbosch}}]{Akushevich:1998ft}%
  \BibitemOpen
  \bibfield  {author} {\bibinfo {author} {\bibfnamefont {I.}~\bibnamefont
  {Akushevich}}, \bibinfo {author} {\bibfnamefont {H.}~\bibnamefont
  {B\"{o}ttcher}}, \ and\ \bibinfo {author} {\bibfnamefont {D.}~\bibnamefont
  {Ryckbosch}},\ }\href@noop {} {\  (\bibinfo {year} {1998})},\ \Eprint
  {http://arxiv.org/abs/hep-ph/9906408} {hep-ph/9906408} \BibitemShut {NoStop}%
\bibitem [{\citenamefont {Brun}\ \emph {et~al.}()\citenamefont {Brun},
  \citenamefont {Hagelberg}, \citenamefont {Hansroul},\ and\ \citenamefont
  {Lassalle}}]{Brun:1978fy}%
  \BibitemOpen
  \bibfield  {author} {\bibinfo {author} {\bibfnamefont {R.}~\bibnamefont
  {Brun}}, \bibinfo {author} {\bibfnamefont {R.}~\bibnamefont {Hagelberg}},
  \bibinfo {author} {\bibfnamefont {M.}~\bibnamefont {Hansroul}}, \ and\
  \bibinfo {author} {\bibfnamefont {J.~C.}\ \bibnamefont {Lassalle}},\
  }\href@noop {} {\ }\bibinfo {note} {CERN-DD-78-2-REV}\BibitemShut {NoStop}%
\bibitem [{\citenamefont {Ravndal}(1973)}]{Ravndal:1973kt}%
  \BibitemOpen
  \bibfield  {author} {\bibinfo {author} {\bibfnamefont {F.}~\bibnamefont
  {Ravndal}},\ }\href {\doibase 10.1016/0370-2693(73)90445-0} {\bibfield
  {journal} {\bibinfo  {journal} {Phys. Lett.}\ }\textbf {\bibinfo {volume}
  {B43}},\ \bibinfo {pages} {301} (\bibinfo {year} {1973})}\BibitemShut
  {NoStop}%
\bibitem [{\citenamefont {Kingsley}(1974)}]{Kingsley:1974vf}%
  \BibitemOpen
  \bibfield  {author} {\bibinfo {author} {\bibfnamefont {R.}~\bibnamefont
  {Kingsley}},\ }\href {\doibase 10.1103/PhysRevD.10.1580} {\bibfield
  {journal} {\bibinfo  {journal} {Phys. Rev.}\ }\textbf {\bibinfo {volume}
  {D10}},\ \bibinfo {pages} {1580} (\bibinfo {year} {1974})}\BibitemShut
  {NoStop}%
\bibitem [{\citenamefont {Cahn}(1978)}]{Cahn:1978se}%
  \BibitemOpen
  \bibfield  {author} {\bibinfo {author} {\bibfnamefont {R.~N.}\ \bibnamefont
  {Cahn}},\ }\href {\doibase 10.1016/0370-2693(78)90020-5} {\bibfield
  {journal} {\bibinfo  {journal} {Phys. Lett.}\ }\textbf {\bibinfo {volume}
  {B78}},\ \bibinfo {pages} {269} (\bibinfo {year} {1978})}\BibitemShut
  {NoStop}%
\bibitem [{\citenamefont {Cahn}\ and\ \citenamefont
  {Jackson}(1990)}]{Cahn:1990jk}%
  \BibitemOpen
  \bibfield  {author} {\bibinfo {author} {\bibfnamefont {R.~N.}\ \bibnamefont
  {Cahn}}\ and\ \bibinfo {author} {\bibfnamefont {J.~D.}\ \bibnamefont
  {Jackson}},\ }\href {\doibase 10.1103/PhysRevD.42.3690} {\bibfield  {journal}
  {\bibinfo  {journal} {Phys. Rev.}\ }\textbf {\bibinfo {volume} {D42}},\
  \bibinfo {pages} {3690} (\bibinfo {year} {1990})}\BibitemShut {NoStop}%
\bibitem [{\citenamefont {Boer}\ and\ \citenamefont
  {Mulders}(1998)}]{Boer:1997nt}%
  \BibitemOpen
  \bibfield  {author} {\bibinfo {author} {\bibfnamefont {D.}~\bibnamefont
  {Boer}}\ and\ \bibinfo {author} {\bibfnamefont {P.~J.}\ \bibnamefont
  {Mulders}},\ }\href {\doibase 10.1103/PhysRevD.57.5780} {\bibfield  {journal}
  {\bibinfo  {journal} {Phys. Rev.}\ }\textbf {\bibinfo {volume} {D57}},\
  \bibinfo {pages} {5780} (\bibinfo {year} {1998})}\BibitemShut {NoStop}%
\bibitem [{\citenamefont {Boer}\ and\ \citenamefont
  {Mulders}(2000)}]{Boer:1999si}%
  \BibitemOpen
  \bibfield  {author} {\bibinfo {author} {\bibfnamefont {D.}~\bibnamefont
  {Boer}}\ and\ \bibinfo {author} {\bibfnamefont {P.~J.}\ \bibnamefont
  {Mulders}},\ }\href {\doibase 10.1016/S0550-3213(99)00719-1} {\bibfield
  {journal} {\bibinfo  {journal} {Nucl. Phys.}\ }\textbf {\bibinfo {volume}
  {B569}},\ \bibinfo {pages} {505} (\bibinfo {year} {2000})}\BibitemShut
  {NoStop}%
\bibitem [{\citenamefont {Rubin}(2009)}]{Rubin:2009zz}%
  \BibitemOpen
  \bibfield  {author} {\bibinfo {author} {\bibfnamefont {J.~G.}\ \bibnamefont
  {Rubin}},\ }\href {\doibase 10.3204/DESY-THESIS-2009-045} {Ph.D. thesis},\
  \bibinfo  {school} {University of Illinois at Urbana-Champaign} (\bibinfo
  {year} {2009}),\ \bibinfo {note} {{DESY-THESIS-2009-045}}\BibitemShut
  {NoStop}%
\bibitem [{\citenamefont {Pumplin}\ \emph {et~al.}(2001)\citenamefont
  {Pumplin}, \citenamefont {Stump}, \citenamefont {Brock}, \citenamefont
  {Casey}, \citenamefont {Huston} \emph {et~al.}}]{Pumplin:2001ct}%
  \BibitemOpen
  \bibfield  {author} {\bibinfo {author} {\bibfnamefont {J.}~\bibnamefont
  {Pumplin}}, \bibinfo {author} {\bibfnamefont {D.}~\bibnamefont {Stump}},
  \bibinfo {author} {\bibfnamefont {R.}~\bibnamefont {Brock}}, \bibinfo
  {author} {\bibfnamefont {D.}~\bibnamefont {Casey}}, \bibinfo {author}
  {\bibfnamefont {J.}~\bibnamefont {Huston}},  \emph {et~al.},\ }\href
  {\doibase 10.1103/PhysRevD.65.014013} {\bibfield  {journal} {\bibinfo
  {journal} {Phys. Rev.}\ }\textbf {\bibinfo {volume} {D65}},\ \bibinfo {pages}
  {014013} (\bibinfo {year} {2001})}\BibitemShut {NoStop}%
\bibitem [{\citenamefont {Pumplin}\ \emph {et~al.}(2002)\citenamefont {Pumplin}
  \emph {et~al.}}]{Pumplin:2002vw}%
  \BibitemOpen
  \bibfield  {author} {\bibinfo {author} {\bibfnamefont {J.}~\bibnamefont
  {Pumplin}} \emph {et~al.},\ }\href {\doibase 10.1088/1126-6708/2002/07/012}
  {\bibfield  {journal} {\bibinfo  {journal} {JHEP}\ }\textbf {\bibinfo
  {volume} {0207}},\ \bibinfo {pages} {012} (\bibinfo {year}
  {2002})}\BibitemShut {NoStop}%
\bibitem [{\citenamefont {{Sj\"ostrand, Torbjorn}}(1994)}]{Sjostrand:1993yb}%
  \BibitemOpen
  \bibfield  {author} {\bibinfo {author} {\bibnamefont {{Sj\"ostrand,
  Torbjorn}}},\ }\href {\doibase 10.1016/0010-4655(94)90132-5} {\bibfield
  {journal} {\bibinfo  {journal} {Comput. Phys. Commun.}\ }\textbf {\bibinfo
  {volume} {82}},\ \bibinfo {pages} {74} (\bibinfo {year} {1994})}\BibitemShut
  {NoStop}%
\bibitem [{\citenamefont {Airapetian}\ \emph {et~al.}()\citenamefont
  {Airapetian} \emph {et~al.}}]{DC19database}%
  \BibitemOpen
  \bibfield  {author} {\bibinfo {author} {\bibfnamefont {A.}~\bibnamefont
  {Airapetian}} \emph {et~al.} (\bibinfo {collaboration} {HERMES
  Collaboration}),\ }\href@noop {} {}\bibinfo {note} {HERMES multiplicity
  database,
  \href{http://www-hermes.desy.de/multiplicities/}{http://www-hermes.desy.de/multiplicities/};
  and Durham HEP database, \href{http://durpdg.dur.ac.uk}
  {http://durpdg.dur.ac.uk}}\BibitemShut {NoStop}%
\end{thebibliography}%

\end{document}